\documentclass[twocolumn, times]{aastex63}
\usepackage[T1]{fontenc}
\usepackage{amsmath}

\received{-}
\revised{-}
\accepted{-}

\submitjournal{AAS Journals}

\shorttitle{TW~Hya}
\shortauthors{Teague et al.}

\graphicspath{{./}{figures/}}

\newcommand{\ms}[1]{$#1~{\rm m\,s^{-1}}$}
\newcommand{\kms}[1]{$#1~{\rm km\,s^{-1}}$}

\newcommand{\mJybeam}[1]{$#1~{\rm mJy\,beam^{-1}}$}

\begin{document}

\title{Mapping the Complex Kinematic Substructure in the TW~Hya Disk}
\shorttitle{Mapping the Complex Kinematic Substructure in the TW~Hya Disk}

\correspondingauthor{Richard Teague}
\email{rteague@mit.edu}

\author[0000-0003-1534-5186]{Richard Teague}
\affiliation{Department of Earth, Atmospheric, and Planetary Sciences, Massachusetts Institute of Technology, Cambridge, MA 02139, USA}
\affiliation{Center for Astrophysics | Harvard \& Smithsonian, 60 Garden Street, Cambridge, MA 02138, USA}

\author[0000-0001-7258-770X]{Jaehan Bae}
\affiliation{Department of Astronomy, University of Florida, Gainesville, FL 32611, USA}

\author[0000-0003-2253-2270]{Sean~M.~Andrews}
\affiliation{Center for Astrophysics | Harvard \& Smithsonian, 60 Garden Street, Cambridge, MA 02138, USA}

\author[0000-0002-7695-7605]{Myriam Benisty}
\affiliation{Univ. Grenoble Alpes, CNRS, IPAG, 38000 Grenoble, France}

\author[0000-0003-4179-6394]{Edwin A. Bergin}
\affil{Department of Astronomy, University of Michigan, 323 West Hall, 1085 S. University Avenue, Ann Arbor, MI 48109, USA}

\author[0000-0003-4689-2684]{Stefano Facchini}
\affiliation{Dipartimento di Fisica, Universit\'{a} degli Studi di Milano, via Celoria 16, 20133 Milano, Italy}

\author[0000-0001-6947-6072]{Jane Huang}
\altaffiliation{NASA Hubble Fellowship Program Sagan Fellow}
\affiliation{Department of Astronomy, University of Michigan, 323 West Hall, 1085 S. University Avenue, Ann Arbor, MI 48109, United States of America}

\author[0000-0003-4663-0318]{Cristiano Longarini}
\affiliation{Dipartimento di Fisica, Universit\'{a} degli Studi di Milano, via Celoria 16, 20133 Milano, Italy}

\author[0000-0003-1526-7587]{David Wilner}
\affiliation{Center for Astrophysics | Harvard \& Smithsonian, 60 Garden Street, Cambridge, MA 02138, USA}

\begin{abstract}
We present ALMA observations of CO $J = 2-1$ and CS $J = 5-4$ emission from the disk around TW~Hydrae. Both molecules trace a predominantly Keplerian velocity structure, although a slowing of the rotation velocity is detected at the outer edge of the disk beyond ${\approx}~140$~au in CO emission. This was attributed to the enhanced pressure support from the gas density taper near the outer edge of the disk. Subtraction of an azimuthally symmetric background velocity structure reveals localized deviations in the gas kinematics traced by each of the molecules. Both CO and CS exhibit a `Doppler flip' feature, centered nearly along the minor axis of the disk (${\rm PA} \sim 60\degr$) at a radius of $1\farcs35$, coinciding with the large gap observed in scattered light and mm~continuum. In addition, the CO emission, both through changes in intensity and its kinematics, traces a tightly wound spiral, previously seen with higher frequency CO $J = 3-2$ observations \citep{Teague_ea_2019a}. Through comparison with linear models of the spiral wakes generated by embedded planets, we interpret these features in the context of interactions with a Saturn-mass planet within the gap at a position angle of ${\rm PA} = 60\degr$, consistent with the theoretical predictions of \citet{Mentiplay_ea_2019}. The lack of a corresponding spiral in the CS emission is attributed to the strong vertical dependence on the buoyancy spirals which are believed to only grow in the atmospheric of the disk, rather than those traced by CS emission.
\end{abstract}


\section{Introduction}

There is mounting evidence that protoplanetary disks exhibit a similar complexity of substructures in their gas density and velocity structures as they do in their dust distributions \citep{DiskDynamics_ea_2020, PPVII_Pinte}. Identifying and characterizing these structures promises to yield unique insights into the planet formation process, as young, recently formed planets can be detected through the subtle dynamical perturbations they impart on the gas through which they move \citep{Perez_ea_2015, Pinte_ea_2018b, Pinte_ea_2019, Casassus_Perez_2019}. Such detections provide opportunities to study planet formation \emph{in action}, and facilitate the detection of planets that evade detection at optical and/or infrared wavelengths. 

Beyond the detection of planets, kinematic-focused observations help localize variations in the rotation speed of the gas, facilitating the inference of the radial gas pressure profile \citep{Teague_ea_2018a, Teague_ea_2018c, Yu_ea_2021} and enabling direct tests of grain trapping in pressure maxima \citep{Rosotti_ea_2020, Yen_Gu_2020, Boehler_ea_2021}. With the Atacama Large Millimeter/submillimeter Array (ALMA) delivering unprecedented spatial resolutions and sensitivities, these analyses have been extended to also probe vertical motions \citep{Teague_ea_2019b, Yu_ea_2021}, revealing large scale flows that are likely examples of meridional flows \citep{Szulagyi_ea_2014, Morbidelli_ea_2014}. Such structures provide a key dynamical link between the chemically rich atmospheric layers of the disk with the relatively inert midplanes, potentially transporting volatile rich gas to accreting planets \citep{Cridland_ea_2020}. Mapping these structures is thus of immense importance in understanding the delivery and formation of exoplanetary atmospheres.

It is clear, therefore, that dedicated studies of the gas velocity structure of disks provide new constraints the key dynamical processes and the presence (or absence) of still-forming planets. Previous studies have focused predominantly on CO emission out of a necessity to achieve a good signal-to-noise ratio across a narrow velocity (or frequency) bandwidth \citep[although there are a few recent exceptions that have been able to use CO isotopologue emission, for example,][]{Teague_ea_2021, Yu_ea_2021}. We present new observations designed specifically for the kinematical analyses of both CO and CS emission from the disk around TW~Hya. The sensitivity of the observations were chosen to demonstrate the utility of investing in longer integrations that allow the study of the dynamics of the gas traced by molecules less abundant than CO, that exists closer to the disk midplane.

As one of the closest protoplanetary disks to Earth \citep[$d = 60.1$~pc;][]{Gaia_ea_2018}, TW~Hya has been extensively studied across a range of frequencies. At sub-mm wavelengths the continuum emission has been mapped at an exquisite ${\sim}$20~mas resolution \citep[1.2~au;][]{Andrews_ea_2016, Macias_ea_2021}, revealing multiple concentric rings. A more recent study, achieving an unprecedented sensitivity, reports the detection of mm continuum emission to larger radii of ${\sim}~100~{\rm au}$ \citep{Ilee_ea_2022}. An intriguing, slightly elongated feature was reported by \citet{Tsukagoshi_ea_2019} which \citet{Zhu_ea_2022} argue could be the spatially resolved envelope of a 10 -- $20~M_{\earth}$ planet. In terms of the gas, CO emission was originally reported by \citet{Zuckerman_ea_1995}, with \citet{Qi_ea_2004} resolving the characteristic emission morphology indicative of Keplerian rotation. At shorter wavelengths, the sub-\micron{} dust in the disk atmosphere has been mapped through scattered light observations from both space- \citep{Roberge_ea_2005, Debes_ea_2013, Debes_ea_2016} and ground-based facilities \citep{Akiyama_ea_2015, Rapson_ea_2015, vanBoekel_ea_2017}. Gaps coincident with those detected in the mm continuum were identified, in addition to a large, broad gap at ${\sim}~90$~au, likely associated with a depletion in CS emission reported by \citet{Teague_ea_2017}, and aligning with the recently reported mm~continuum gap \citep{Ilee_ea_2022}. A kinematic analysis of the high spatial resolution, but low spectral resolution, images of CO emission presented in \citet{Huang_ea_2018} found a tightly wound spiral feature in the gas velocity structure on top of the background rotation \citep{Teague_ea_2019a}. Such spiral features were proposed by \citet{Bae_ea_2021} to be buoyancy resonances driven by an embedded planet within the outer gap.

As such, TW~Hya represents an ideal candidate for a thorough kinematic analysis to fully characterize the possibilities of on-going planet-disk interactions. In Section~\ref{sec:observations} the observations and the subsequent calibration and imaging are described, including the generation of moment maps. A study of the velocity structure is performed in Section~\ref{sec:velocities}, looking at both bulk background motion, and localized deviations and deprojecting disk-frame velocities. The interpretation of these velocity structures are discussed in Section~\ref{sec:discussion}, with the main findings summarized in Section~\ref{sec:summary}.

\section{Observations}
\label{sec:observations}

\begin{figure*}
    \centering
    \includegraphics[width=\textwidth]{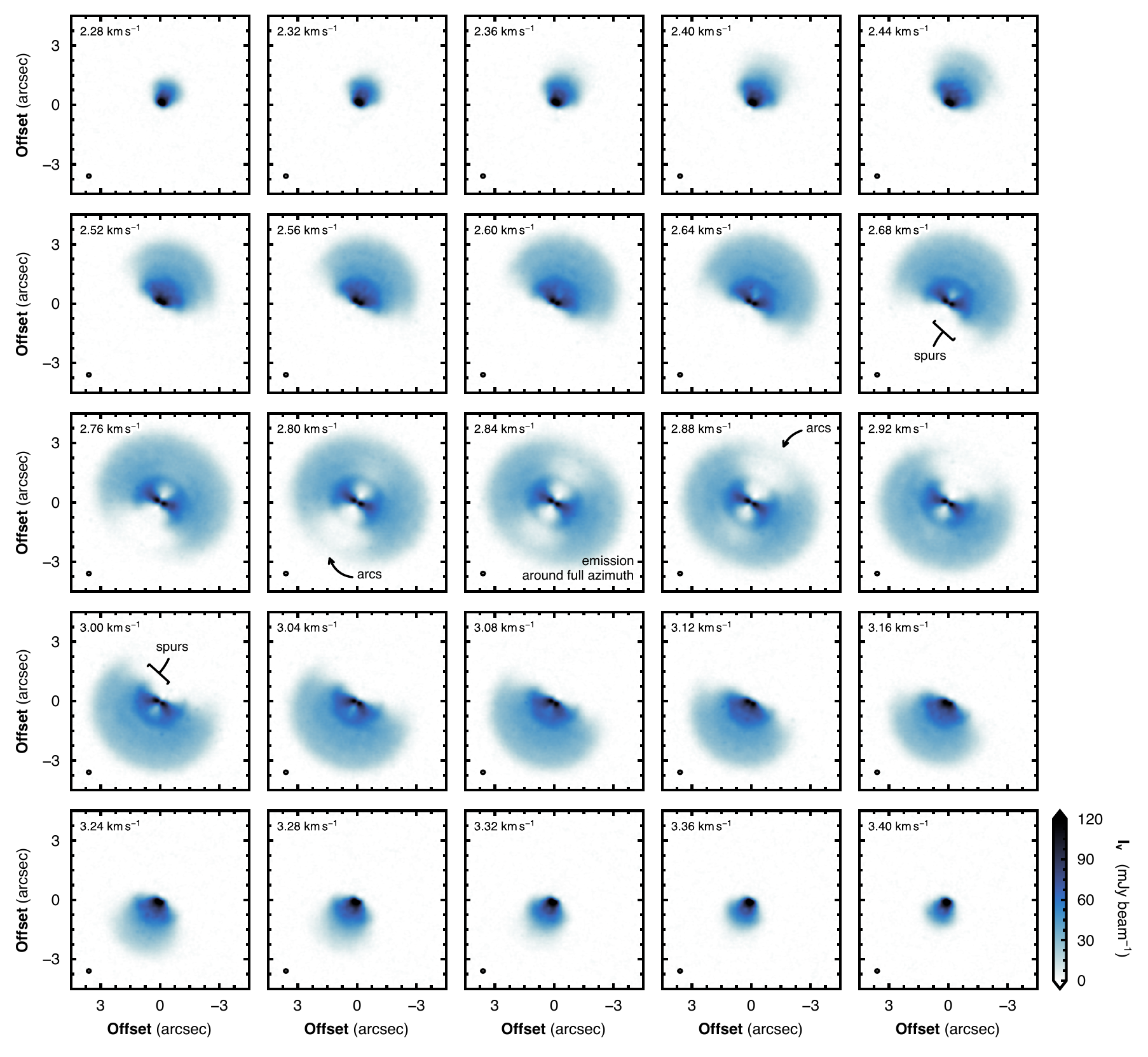}
    \caption{CO (2--1) channel maps, imaged at \ms{40} spacing. The beam size is shown in the bottom left of the panels. The channel velocity is shown in the top left of each panel. Each panel spans $\pm 4.5\arcsec$ ($\pm 270$~au) across. Note that the color scaling has been adjusted to bring out the detail in the line wings and the peak intensity is \mJybeam{176.4}.}
    \label{fig:channels_CO_40ms}
\end{figure*}

Observations were acquired as part of project 2018.A.00021.S (PI: Teague). \citet{Teague_ea_2022} described the observational setup and calibration of these data in detail. We also combined this with data from project 2013.1.00387.S (PI: Guilloteau), originally presented in \citet{Teague_ea_2016}, which has an almost identical correlator set-up. In this paper we focus on the CO and CS emission.

\subsection{Post Processing}

We start with the calibrated measurement sets from \citet[`2013 data']{Teague_ea_2016} and \citet[`2018' data]{Teague_ea_2022}. The continuum fluxes were compared using the tools\footnote{\url{https://bulk.cv.nrao.edu/almadata/lp/DSHARP/scripts/reduction_utils.py}} provided by the DSHARP Large Program \citep{Andrews_ea_2018}. The two continuum fluxes were found to be within $\lesssim~2\%$ of one another so no flux rescaling was applied. As the 2013 data was already aligned, the phase center was updated to match that of the 2018 data using the \texttt{fixplanets} command in CASA to account for any proper motion of the source.

\subsection{Imaging}

As in \citet{Teague_ea_2022}, Briggs weighting with a robust value of 0.5 and a channel width of \ms{40} was adopted for the imaging of both CO and CS emission. This resulted in a synthesized beam of $0\farcs19 \times 0\farcs17$ ($89\degr$) and $0\farcs18 \times 0\farcs17$ ($87\degr$) for CO and CS, respectively. A Keplerian mask was generated for both molecules for the \texttt{CLEAN}ing,\footnote{\url{github.com/richteague/keplerian\_mask}} making sure that all disk emission was contained within the mask. Following \citet{Teague_ea_2018c}, a stellar mass of $M_{\star} = 0.6~M_{\odot}$ and a viewing geometry described by $i = 5\fdg8$ and ${\rm PA} = 151\degr$ were adopted for this mask. After the imaging, a correction was applied to the image to account for the non-Gaussian synthesized beams due to the combination of several different array configurations \citep[see the discussion in][]{Czekala_ea_2021, Casassus_Carcamo_2022}, as proposed by \citet{Jorsater_vanMoorsel_1995}. The resulting RMS in a line free channel was measured to be \mJybeam{0.6} and \mJybeam{0.7} for CO and CS, respectively. Integrating over the Keplerian mask for CO and CS, we recover integrated intensities of $18.31 \pm 0.03~{\rm Jy~km\,s^{-1}}$ and $1.69 \pm 0.03~{\rm Jy~km\,s^{-1}}$. These uncertainties do not include a systematic uncertainty of ${\sim}~10\%$ associated with the flux calibration of the data.

\subsection{Channel Maps}
\label{sec:observations:channel_maps}

\begin{figure*}
    \centering
    \includegraphics[width=\textwidth]{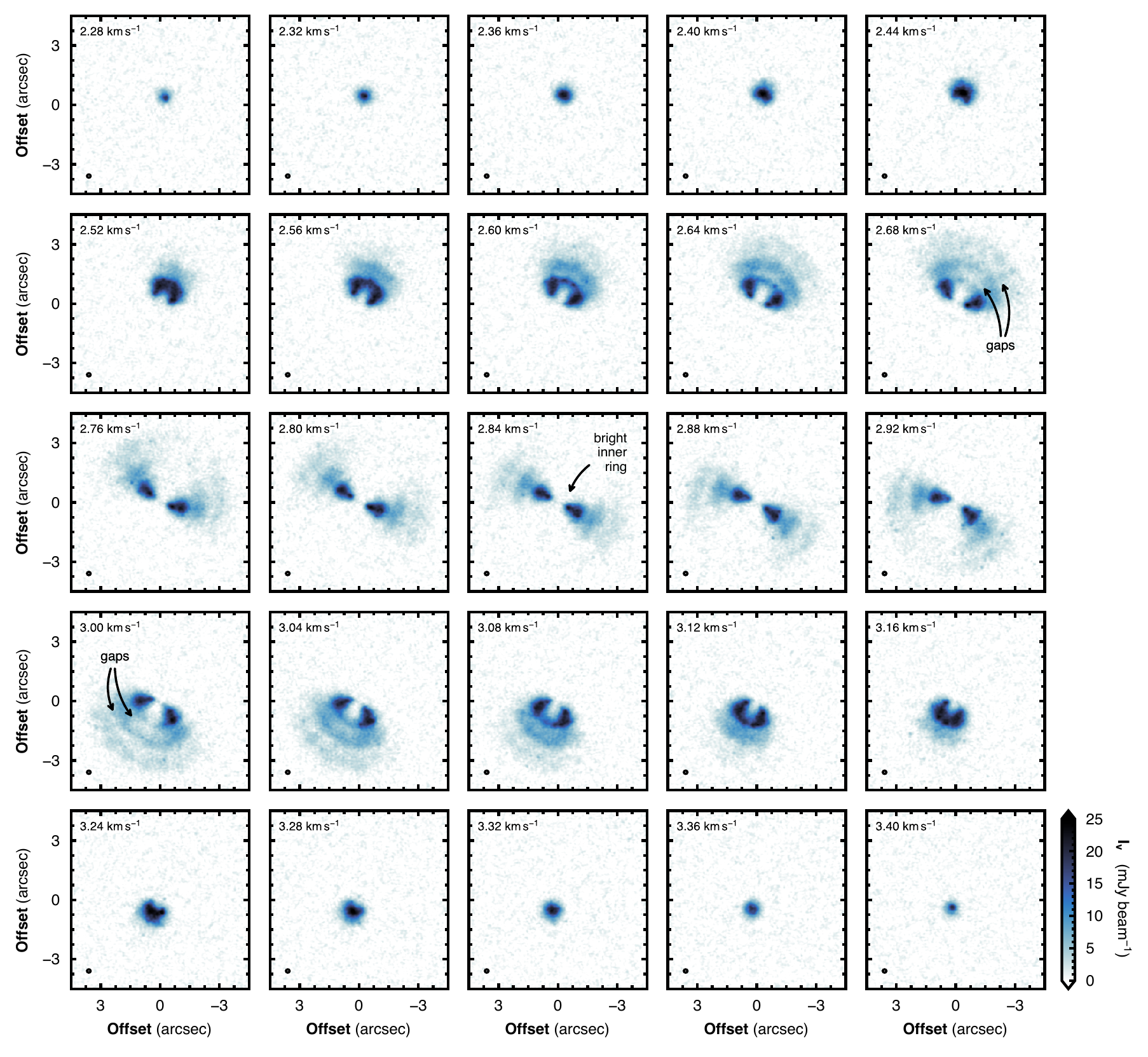}
    \caption{CS (5--4) channel maps, imaged at \ms{40} spacing. The beam size is shown in the bottom left of the panels and the channel velocity is shown in the top left of each panel. Each panel spans $\pm 4.5\arcsec$ ($\pm 270$~au) across. Note that the color scaling has been adjusted to bring out the detail in the line wings. The peak intensity is \mJybeam{26.4}.}
    \label{fig:channels_CS_40ms}
\end{figure*}

Channel maps from the images are presented in Figures~\ref{fig:channels_CO_40ms} and \ref{fig:channels_CS_40ms} for CO, and CS, respectively. Only the central channels where there is extended emission are shown, although emission close to the disk center is detected out to velocities of $\pm~{\sim}3~{\rm km\,s^{-1}}$ relative to the systemic velocity of \kms{2.84} for CO \citep[comparable to that reported by][]{Rosenfeld_ea_2012} and $\pm~{\sim} 1~{\rm km\,s^{-1}}$ for CS. Both lines exhibit a high level of substructure that is seen across multiple channels. The CO emission morphology is dominated by large arcs such that in the central channels (those close to the systemic velocity of \kms{2.84}) emission can be seen extending around the full azimuth of the disk. In these central channels two rings are seen bridging the two lobes of emission across the major axis of the disk suggesting regions of higher density. These arcs can also be seen in channels offset from the line center by ${\sim}$\ms{100} as small protrusions from the bulk emission.

The CS emission shows a similar level of substructure, but with a different morphology, likely because the emission is optically thin and is thus more sensitive to changes in the background gas density or abundance structure than CO emission. The inner regions are dominated by a bright ring, out to a radius of ${\sim}~1\farcs5$, with a more diffuse outer disk. Two gaps are observed at ${\sim}~1\farcs5$ \citep[originally reported in][]{Teague_ea_2017}, and ${\sim}~2\farcs6$ \citep[156~au; previously reported by][]{Teague_ea_2018b}, which can be traced in many channels around the systemic velocity.

\subsection{Moment Maps}
\label{sec:observations:moment_maps}

\begin{figure*}
    \centering
    \includegraphics[width=\textwidth]{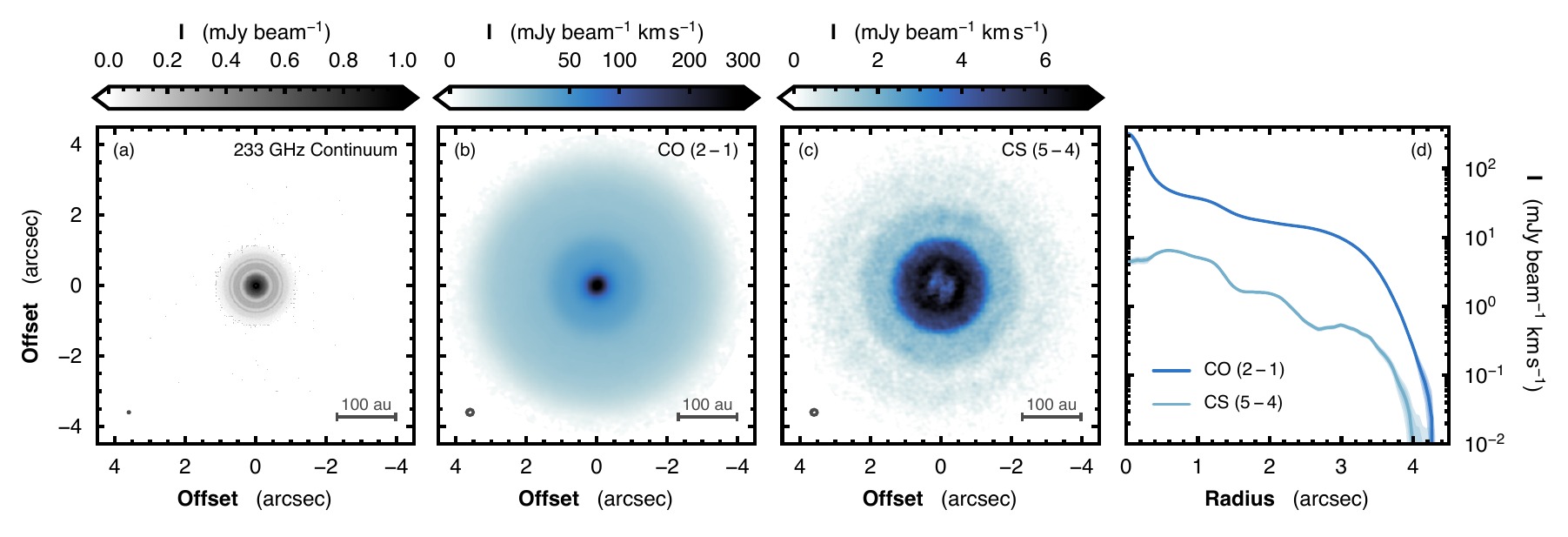}
    \caption{Comparison of the archival 233~GHz continuum from \citet{Macias_ea_2021}, (a), and the velocity-integrated intensity (zeroth moment) maps of CO, (b) and CS, (c). Note that the CO data has a `square-root' stretch applied in order to bring out the morphology. In each panel the synthesized beam is shown in the lower left and a 100~au scale bar in the lower right. The right-most panel, (d), show the azimuthally averaged radial profiles of the CO and CS emission. The shaded regions show uncertainties in steps of $1\sigma$.}
    \label{fig:continuum_moment_zero_maps}
\end{figure*}

To aid in the analysis of the data we collapse all data along the spectral axis into 2D maps using the methods described below. All collapsing was done with the \texttt{bettermoments}\footnote{\url{github.com/richteague/bettermoments}} Python package.

\subsubsection{Integrated Intensity}

For the total integrated intensity, or zeroth moment maps, we integrated the emission along the spectral axis using the Keplerian \texttt{CLEAN} mask to circumvent any sigma clipping. The CO zeroth moment map has already been published in \citet{Teague_ea_2022} where a shadow was detected in the outer disk. These moment maps are shown in Fig.~\ref{fig:continuum_moment_zero_maps} and compared to the 233~GHz continuum originally presented in \citet{Macias_ea_2021} to highlight the difference in the radial extent of the molecular gas and the mm-sized grains. We use the Python package \texttt{GoFish} \citep{GoFish} to produce azimuthally averaged radial profiles of these zero moment maps. Emission from CO is detected out to $4\farcs2$ (252~au) and CS is detected out to $3\farcs9$ (235~au), while the continuum edge is around $1\farcs65$ \citep[100~au;][]{Ilee_ea_2022}. In these radial profiles, the substructures identified in the CO and CS channel maps are more apparent, most notably the inner cavity and two gaps in the CS emission at $\approx 1\farcs5$ (90~au) and $\approx 2\farcs6$ (156~au).

\subsubsection{Analytical Spectral Profiles}

To calculate maps of line profile properties, such as the line center, width and peak intensity, fitting of analytical profiles with \texttt{bettermoments} was found to yield the best results owing to the high spectral resolution of the data. This is in contrast to the results presented in \citet{Teague_ea_2019a}, where a quadratic curve was fit to the velocity pixel of peak intensity and its two adjacent velocity pixels \citep[see, e.g.,][]{Teague_Foreman-Mackey_2018}, as the data was taken at a much lower spectral resolution. At higher spectral resolutions the quadratic method is limited by the low level of curvature at the line center for a well resolved Gaussian profile. This was particularly an issue for the CO emission which was found to have a highly saturated core due to the large optical depths \citep[as discussed in][]{Teague_ea_2016}.

While the CS was well reproduced by a Gaussian profile,
\begin{equation}
    I(v) = I_0 \exp\left( -\left( \frac{v - v_0}{\Delta V} \right)^2 \right),
\end{equation}
with $v_0$ describing the line center, $\Delta V$ is the Doppler width of the line (where FWHM $= 2 \sqrt{{\rm ln}2}~\Delta V$) and $I_0$ is the peak intensity, the CO emission profile was modeled as,
\begin{equation}
    I(v) = I_0 \cdot \frac{1 - \exp\big(-\tau(v)\big)}{1 - \exp(-\tau_0)},
\end{equation}
with the optical depth $\tau(v)$ varying as a Gaussian with a width of $\Delta V$ and a peak optical depth of $\tau_0$. Note that $\tau_0$ should not be taken as an absolute measure of the column density, but rather a measure of how saturated the core of the line -- which is related to the optical depth of the line -- is with larger values relating to more saturated cores. It was tested whether a Gauss-Hermite quadrature expansion, as is often used for studies of galactic dynamics \citep[e.g.,][]{vanderMarel_Franx_1993}, provided a better fit than an `optically thick Gaussian' to the CO emission. However the Gauss-Hermite expansion failed to describe the data as well as the `optically thick Gaussian' in addition to requiring an additional free parameter.

A straight forward minimization of $\chi^2$ was achieved with the \texttt{curve\_fit} routine in \texttt{scipy} \citep{scipy} to yield an initial fit to the data. Although fast, this approach often found local minima in the $\chi^2$ surface resulting in moment maps with a large variance over small spatial ranges. Instead, an MCMC sampling approach, implemented with the \texttt{emcee} package \citep{emcee}, was found to be more forgiving in such cases. Taking the median of the posterior distributions as the `best-fit' parameter produced far more spatially-coherent maps. For this fitting the Keplerian \texttt{CLEAN} masks were adopted and only pixels with at least as many unmasked voxels as there were free parameters in the model (4 for the CO emission, 3 for the CS) were included in the fit. It was tested whether smoothing with a simple top-hat kernel with a width of 4 channels (\ms{120}) would improve the results, but little difference was found between maps made with or without the smoothing. For this manuscript we focus primarily on $v_0$, shown in Fig.~\ref{fig:v0maps}, while the moment maps for each additional parameter can be found in Appendix~\ref{sec:app:moments}. An exploration of how accurately the parameters are recovered in the presence of noise can be found in Appendix~\ref{sec:app:accuracy}.

\section{Mapping the Velocity Structure}
\label{sec:velocities}

In this Section we explore how we can decompose the observed line-of-sight velocity into its constituent parts in order to map the velocity structure of the disk and isolate localized velocity perturbations.

\subsection{Background Motions}
\label{sec:decomposition:background}

\begin{figure}
    \includegraphics[width=\columnwidth]{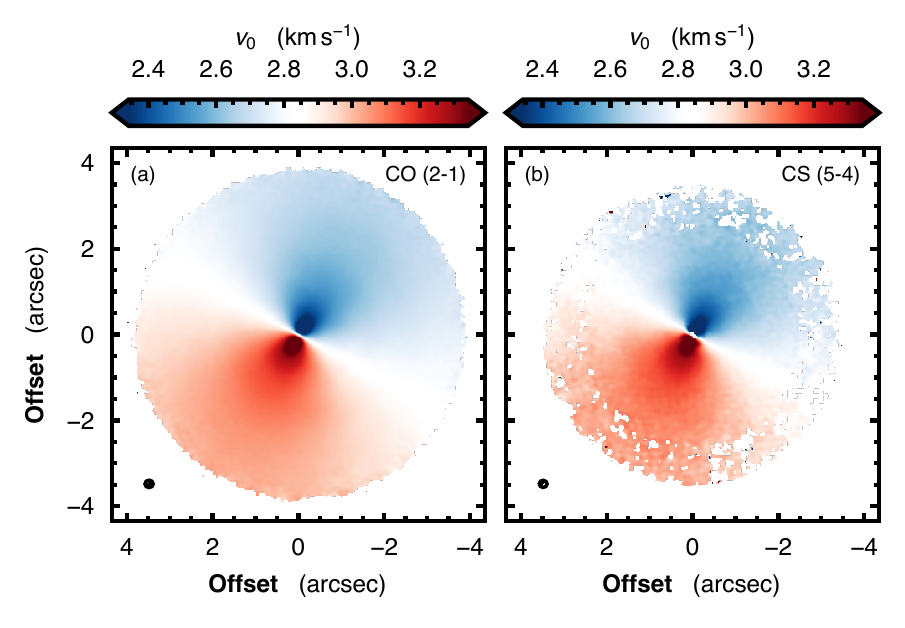}
    \caption{Line of sight velocity maps, $v_0$, for CO, left, and CS, right. The synthesized beam is shown in the bottom left of each panel. The $v_0$ maps are masked where the peak intensities fall below a $5\sigma$ level for CO and $3\sigma$ level for CS.}
    \label{fig:v0maps}
\end{figure}

It is possible to model the background rotation and subtract this component to reveal any non-Keplerian motions within the disk \citep[e.g.,][]{Casassus_Perez_2019, Casassus_ea_2021, Teague_ea_2019a, Woelfer_ea_2021}. There are multiple approaches that can be employed for such a goal, each with their own benefits and limitations. One main consideration is whether to try and model the disk as a whole, thereby imposing some functional form to $v_{\phi}$, such as a purely Keplerian rotation profile \citep[e.g.,][]{Teague_ea_2019a, Teague_ea_2021, Rosotti_ea_2020, Woelfer_ea_2021, Izquierdo_ea_2021, Izquierdo_ea_2022}, or splitting the disk into concentric annuli and fitting each annulus independently, allowing for a far more flexible profile \citep[e.g.,][]{Teague_ea_2018a, Teague_ea_2018c, Teague_ea_2019b, Casassus_Perez_2019, Casassus_ea_2021}.

One consideration is that with improvements in the precision at which rotation curves can be measured, systematic deviations associated with the radial pressure profile \citep{Dullemond_ea_2020, Teague_ea_2021} or disk self-gravity \citep{Veronesi_ea_2021} are becoming apparent, resulting in large residuals at the outer edge of the disk. This issue can be circumvented with the independent annuli approach. However, as a model comprised of independent annuli is highly flexible it can run the risk of over-fitting the data and confusing the disentangling of various velocity components, particularly if the background disk is not azimuthally symmetric \citep[see the discussion regarding the choice of a background velocity profile in][]{Teague_ea_2019b}. As there is substantial structure in the TW~Hya disk that is not azimuthally symmetric, we focus here on just the analytical models for $v_{\phi}$.

\begin{figure*}
    \centering
    \includegraphics[width=\textwidth]{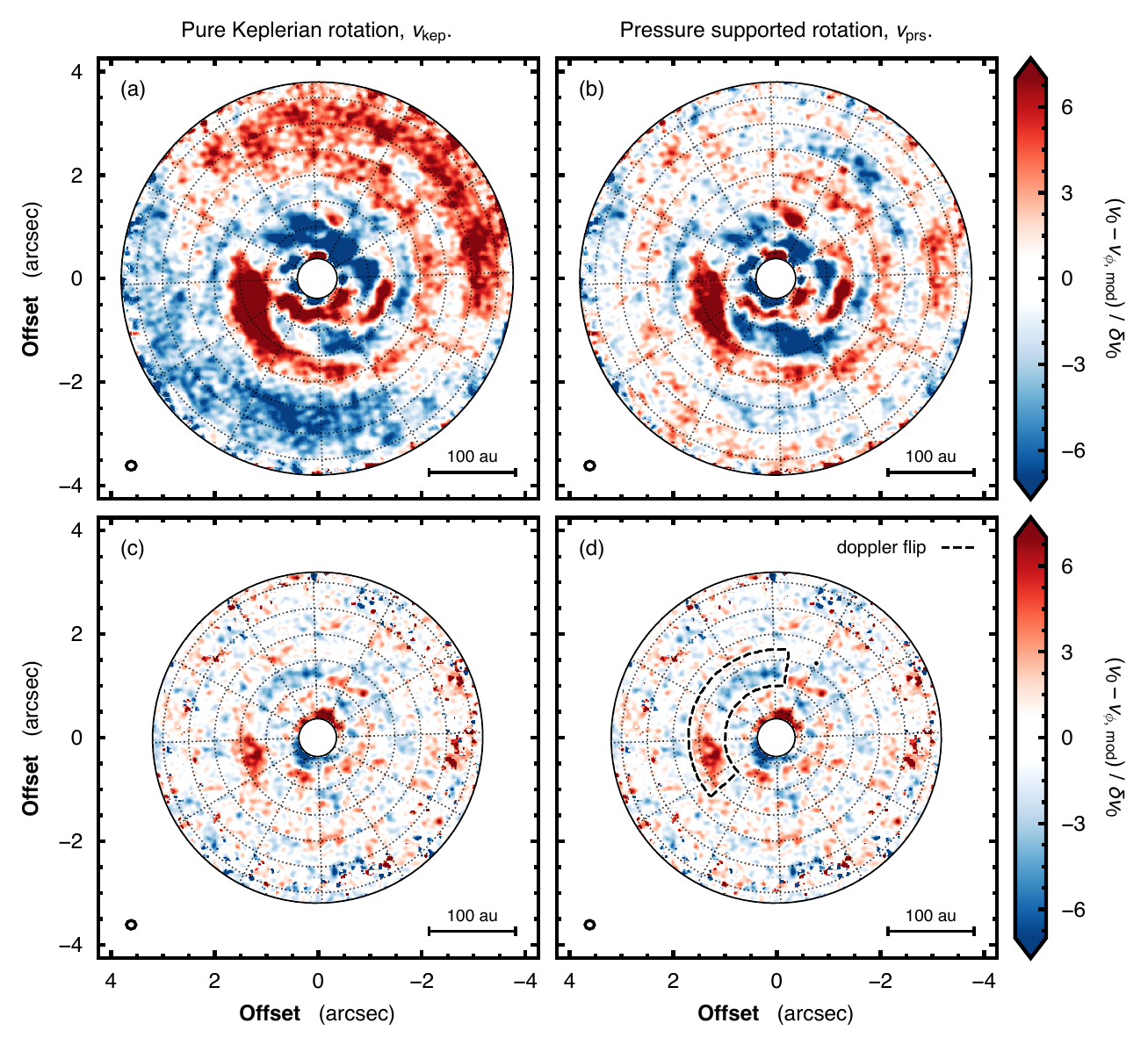}
    \caption{The $v_0$ residuals for CO, top, and CS, bottom, after subtracting a purely Keplerian background model, $v_{\rm kep}$, left column, or a pressure-supported Keplerian model, $v_{\rm prs}$, right column. In each panel the synthesized beam is shown in the lower left, while a 100~au scale bar is shown in the lower right. The dotted contours show radial steps of $0\farcs5$ and azimuthal steps of $30\degr$. The colorbar shows the residuals in units of $\sigma$, with absolute residuals less than $1\sigma$ left white. The dashed contour in (d) highlights the Doppler flip.}
    \label{fig:v0_residuals}
\end{figure*}

\subsubsection{Analytical Models}
\label{sec:decomposition:background:analytical}

To perform the fitting of the background rotation profile to the $v_0$ maps we use the \texttt{eddy} Python package \citep{eddy} and fit a Keplerian rotation profile of the form,
\begin{equation}
    v_{\rm kep} = \sqrt{\frac{GM_{\star}r^2}{(r^2 + z^2)^{\frac{3}{2}}}},
    \label{eq:vkep}
\end{equation}
where $M_{\star}$ is the dynamical stellar mass and $(r,\,z)$ are the disk-frame cylindrical radius and height, related to the on-sky pixel through the disk inclination, $i$, position angle, PA, and a source center offset, $(x_0,\, y_0)$. Thus for this model a total of 5 free parameters are needed: $\Theta_{\rm kep} = \{x_0,\, y_0,\, {\rm PA},\, M_{\star},\, v_{\rm LSR}\}$. For face-on disks there is an extreme degeneracy between $M_{\star}$ and $i$, so we fix $i = 5\fdg8$ following \citet{Teague_ea_2019a}. We note that this is different to other recent works studying TW Hya, such as \citet{Huang_ea_2018} and \citet{Macias_ea_2021}, which adopt a lower $i = 5\fdg0$, however we adopt the higher inclination for a more direct comparison of the kinematical features described in \citet{Teague_ea_2019a}. We have verified that at such low inclinations, the structures and features described in this manuscript are all robust to the choice of inclination.

As previously mentioned, there is mounting evidence that many disks exhibit non-Keplerian rotation due to the contributions of disk self-gravity \citep{Veronesi_ea_2021} or the radial pressure gradient \citep{Dullemond_ea_2020, Teague_ea_2021}. To mimic the effect of these on the rotation curve we have included a pseudo disk mass term, $M_{\rm disk}$, where positive values hasten the rotation speed and negative values slow the rotation. For the case of a sub-Keplerian outer disk, three free parameters are included to the modeling: $\Theta_{\rm prs} = \{\Theta_{\rm kep},\, M_{\rm disk},\, \gamma,\, r_{\rm in}\}$. $r_{\rm in}$ describes the cylindrical radius where $v_{\phi}$ begins to deviate from Keplerian rotation, while $M_{\rm disk}$ dictates the magnitude of this deviation at the outer edge of the disk (for this fitting we set $r_{\rm out} = 4\arcsec$). In essence, $v_{\phi}$ is a Keplerian profile around a star with a dynamical mass of $M_*$ inside $r_{\rm in}$, and approaches a Keplerian profile around a star with a dynamical mass of $M_* + M_{\rm disk}$ (remembering that $M_{\rm disk}$ is a \emph{negative} value in this case) at $r_{\rm out}$. How quickly the effective dynamical mass of the system changes with radius is governed by $\gamma$, with $\gamma = 0$ resulting in a linear progression and $\gamma > 0$ describing a quick initial change which then slowly approaches the final value. The full parameterization used in the \texttt{eddy} fitting is described in more detail in Appendix~\ref{sec:app:beyond_keplerian}. Plots of how the $v_{\phi}$ profile changes under various choices of $\Theta_{\rm prs}$ are also shown.

For each map we consider two models: pure Keplerian rotation, $v_{\rm kep}$, requiring 5 free parameters, $\Theta_{\rm kep}$, and pressure support rotation, $v_{\rm prs}$, requiring 8 free parameters, $\Theta_{\rm prs}$. We model the posterior distributions through \texttt{eddy} which wraps the \texttt{emcee} MCMC sampling package \citep{emcee}. Only regions between $0\farcs4$ (roughly twice the beam FWHM) and out to $3\farcs75$ for CO and $3\farcs5$ for CS, were fit, where the inner boundary was to chosen to avoid effects from beam smearing, while the outer boundary removed regions with low SNR measurements of $v_0$. The cubes were spatially down-sampled such that only spatially independent pixels were considered. 128 walkers were initialized in a cluster around previous literature values, and then given 5,000 steps to burn in and an additional 5,000 steps to sample the posterior distribution. We find consistent posterior distributions for all parameters between both lines and for those from previous works \citep{Huang_ea_2018, Teague_ea_2019b}. Neglecting the nuisance source-center offset parameters, the median values for the Keplerian model were found to be ${\rm PA} = 151\fdg6$, $M_{\star} = 0.82~M_{\odot}$ and $v_{\rm LSR} =$~\ms{2843}. The width of the posteriors (typically used as a measure of the statistical uncertainty) were exceptionally narrow resulting in fractional uncertainties of $\lesssim 0.1\%$ for all parameters. Such a low statistical uncertainty is likely a symptom of an relatively inflexible model rather than a realistic constraint on those parameters and as such are not quoted to avoid misrepresentation.

For the CO the outer ${\sim}~1\arcsec$ was found to be over-predicted with a Keplerian model suggesting a sub-Keplerian outer disk. With the pressure support rotation model Gaussian posteriors were found for the three additional parameters with $M_{\rm disk} = -0.12$, $\gamma = 1.44$ and $r_{\rm in} = 1\farcs71$, again with the widths of the posteriors being below 1\% of these quoted values. The five other parameters were consistent with the values found for the purely Keplerian fit, including the dynamical mass of the star. No similar constraints could be made for the $v_0$ map calculated from CS emission, likely due to the low SNR of the emission in the regions where the deviation is strongest \textbf{($r \gtrsim 3\arcsec$)}.

\subsection{Non-Keplerian Motions}
\label{sec:decomposition:non-keplerian}

\begin{figure*}
    \centering
    \includegraphics[width=\textwidth]{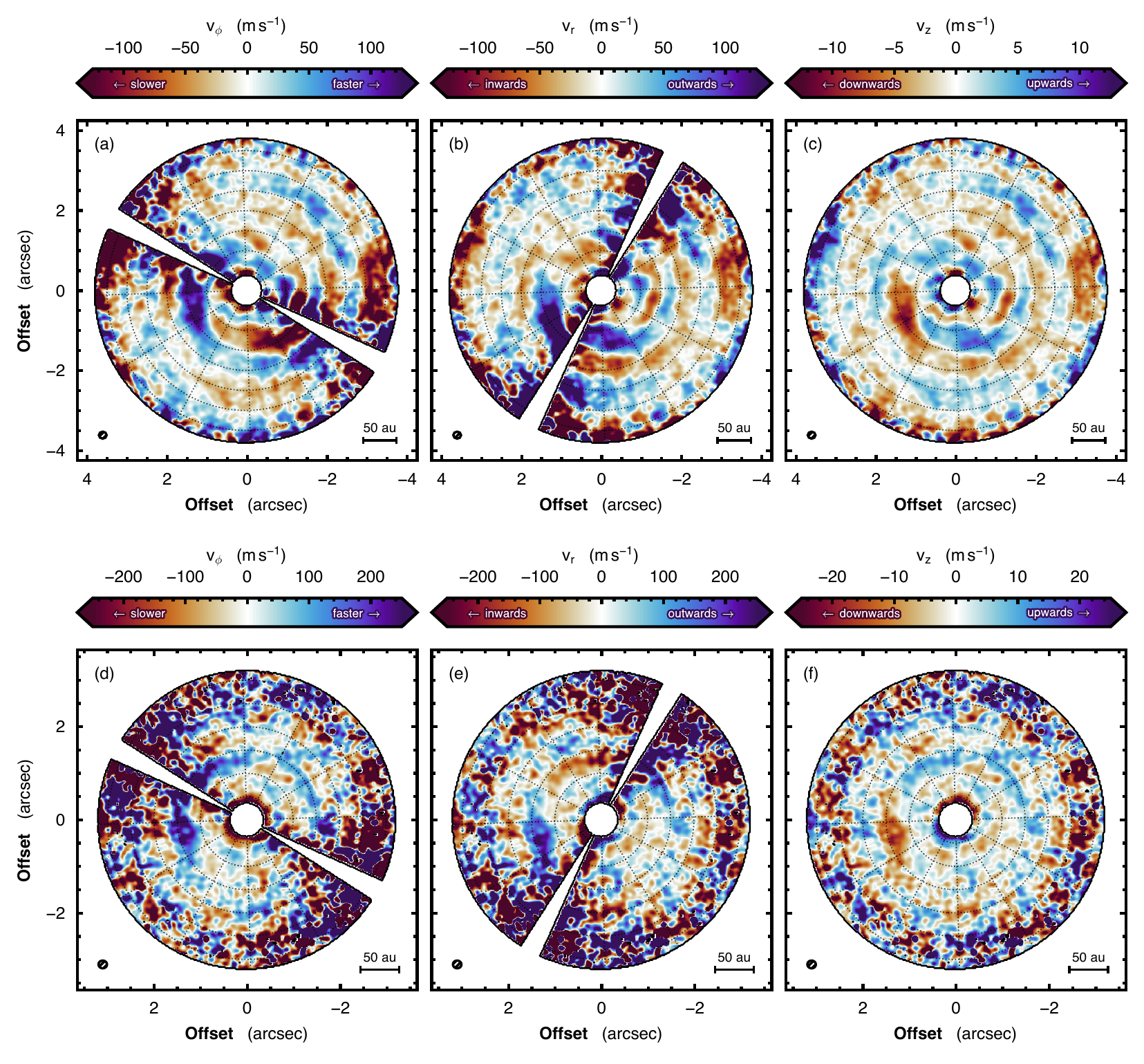}
    \caption{The deprojected velocity residuals for CO, top row, and CS, bottom row, assuming all velocities are azimuthal, left column, radial, central column, or vertical, right column. For the azimuthal and radial velocities wedges along the minor and major axes have been masked as here the observations are insensitive to those components. In each panel the synthesized beam and a 50~au scale bar are shown in the bottom left and right corners. Note that the field of view for the bottom row is smaller than the top row.}
    \label{fig:deprojected_v0_residuals}
\end{figure*}

Figure~\ref{fig:v0_residuals} shows the residuals after subtracting the background velocity structures inferred in the previous subsection. Panels (a) and (b) compare the CO velocity residuals after subtracting a purely Keplerian profile, $v_{\rm kep}$, and a profile including a sub-Keplerian outer disk, $v_{\rm prs}$, respectively. The sub-Keplerian outer edge is clear in panel (a) with negative and positive residuals along the red-shifted and blue-shifted major axis of the disk which are mostly removed when allowing for the more flexible $v_{\phi}$ profile. This asymmetry is removed with the pressure supported model in (b), with little structure seen in the outer regions of the disk. No difference can be seen between the CS residuals when changing between $v_{\rm kep}$ and $v_{\rm prs}$ as the background model. This is as expected as the model parameters for the pressure support were unable to converge.

The most striking feature is a large, tightly wound spiral positive residual, starting at roughly $(1\farcs5,\, 0\farcs0)$, and extending out to the disk edge after almost a full $2\pi$ winding. The start of this spiral was previously detected by \citet{Teague_ea_2019a} using CO $(3-2)$ emission, but due to the lower spectral resolution of that data, was only considered an arc, rather than a full spiral. At smaller radii, two arcs of positive residuals are seen: one directly south of the disk center at an offset of ${\sim}~0\farcs6$, and a second in a south-westerly offset from the star at an offset of ${\sim}~1\arcsec$. Both these features were seen in the CO $(3-2)$ data, suggesting that these are real features, and that these observations are hitting the kinematical equivalent of the `confusion limit'.

The velocity structure traced by CS emission shows far less structure due to the lower SNR of the emission. However, a large positive feature is detected coincident with the peak residual near the start of the spiral arm detected in CO. A weaker negative residual is seen ahead (the disk is rotating in a clockwise direction), up to a ${\rm PA} \sim 0\degr$, before turning into a positive residual again. Such a feature is highly reminiscent of the `Doppler flip' found in HD~100546 \citep{Casassus_Perez_2019}, proposed to be due to the spiral shocks from an embedded planet \citep[e.g.,][]{Perez_ea_2015}. This feature is marked in Fig.~\ref{fig:v0_residuals}d with a dashed contour. Guided by the presence of this feature in the CS map, a similar feature is potentially traced by the CO emission at the same location.

No obvious kinematic deviations are observed around the continuum feature reported by \citet{Tsukagoshi_ea_2019} close to the south-eastern minor axis of the disk (${\rm PA}~{\sim}~237\degr$) at a distance of 52~au ($0\farcs87$).

\subsection{Deprojecting Velocity Components}
\label{sec:decomposition:deprojection}

The measured lines of sight velocity shown in Fig.~\ref{fig:v0maps} is the sum of the projection of all local velocity components,
\begin{equation}
    v_0 = v_{\rm LSR} + v_{\phi,\, {\rm proj}} + v_{r,\, {\rm proj}} + v_{z,\, {\rm proj}},
    \label{eq:v0}
\end{equation}
where the projection of each component along the line of sight is given by,
\begin{subequations}
\begin{align}
    v_{\phi,\, {\rm proj}}  &= +v_{\phi} \cos(\phi) \sin(|i|),\label{eq:vphi_proj}\\
    v_{r,\, {\rm proj}}     &= -v_{r} \sin(\phi) \sin(i),\label{eq:vr_proj}\\
    v_{z,\, {\rm proj}}     &= -v_{z} \cos(i),\label{eq:vz_proj}
\end{align}
\end{subequations}
where $\phi$ is the polar angle in the disk, defined to be $\phi = 0\degr$ along the red-shifted major axis, and increasing in an anti-clockwise direction (i.e., increasing with PA). As the projection of each velocity component has a different azimuthal dependence, it is possible to leverage this difference to disentangle the velocity residuals. For example, if a feature with strong $v_{\phi}$ components crosses the minor axis of the disk, $\phi = \pm 90\degr$, then the projected velocity $v_{\phi,\, {\rm proj}}$, which is observed, will flip sign. Conversely, for a $v_r$ component that crosses the minor axis of the disk will not flip sign; the observed $v_{r,\, {\rm proj}}$ will remain constant. Therefore, under the assumption that the velocity residuals are driven by a single velocity component, we can deproject the residuals in Fig.~\ref{fig:v0_residuals} to recover the true, intrinsic velocity structure. If these are found to exhibit multiple changes in sign coincident with disk axes, an unlikely physical situation, then a strong argument can be made against that velocity component being responsible for the observed residuals. 

To apply this technique we deprojected the residuals after subtracting the best pressure-supported Keplerian rotation model assuming that the perturbations were fully azimuthal, radial or vertical in nature through Eqns.~\ref{eq:vphi_proj}, \ref{eq:vr_proj} and \ref{eq:vz_proj}, respectively. Figure~\ref{fig:deprojected_v0_residuals} shows the results for CO in the top row, and CS in the bottom row. For CO it is clear that a majority of the structure cannot be azimuthal or radial in nature due to the numerous changes in sign observed across both axes. Indeed, the azimuthal coherence of the spiral-shaped residuals point towards to a vertical component, with gas moving from the atmosphere towards the disk midplane, as argued for in \citet{Teague_ea_2019a}.

Around the north-east disk minor axis (${\rm PA} \sim 60\degr$), the interpretation is less straightforward. Both CO and CS show residuals that under the assumption of rotational velocity components yield the most azimuthally coherent features, as shown in Figs.~\ref{fig:deprojected_v0_residuals}b and d. A region of hastened rotation is expected at the outer edge of a gap due to the large positive radial pressure gradient, as would be expected based on the substantial surface density perturbation at that location \citep{Teague_ea_2017, vanBoekel_ea_2017}. However, such a feature would be expected to be azimuthally symmetric and not limited to just the north-east half of the disk. In this scenario, one could argue that the large-scale spiral traced by CO may hide the super-Keplerian rotating gas along the south-east side of the disk, however the lack of a similar spiral feature traced by the CS emission would therefore require a different explanation for the deeper regions of the disk.

\begin{figure}
    \centering
    \includegraphics[width=\columnwidth]{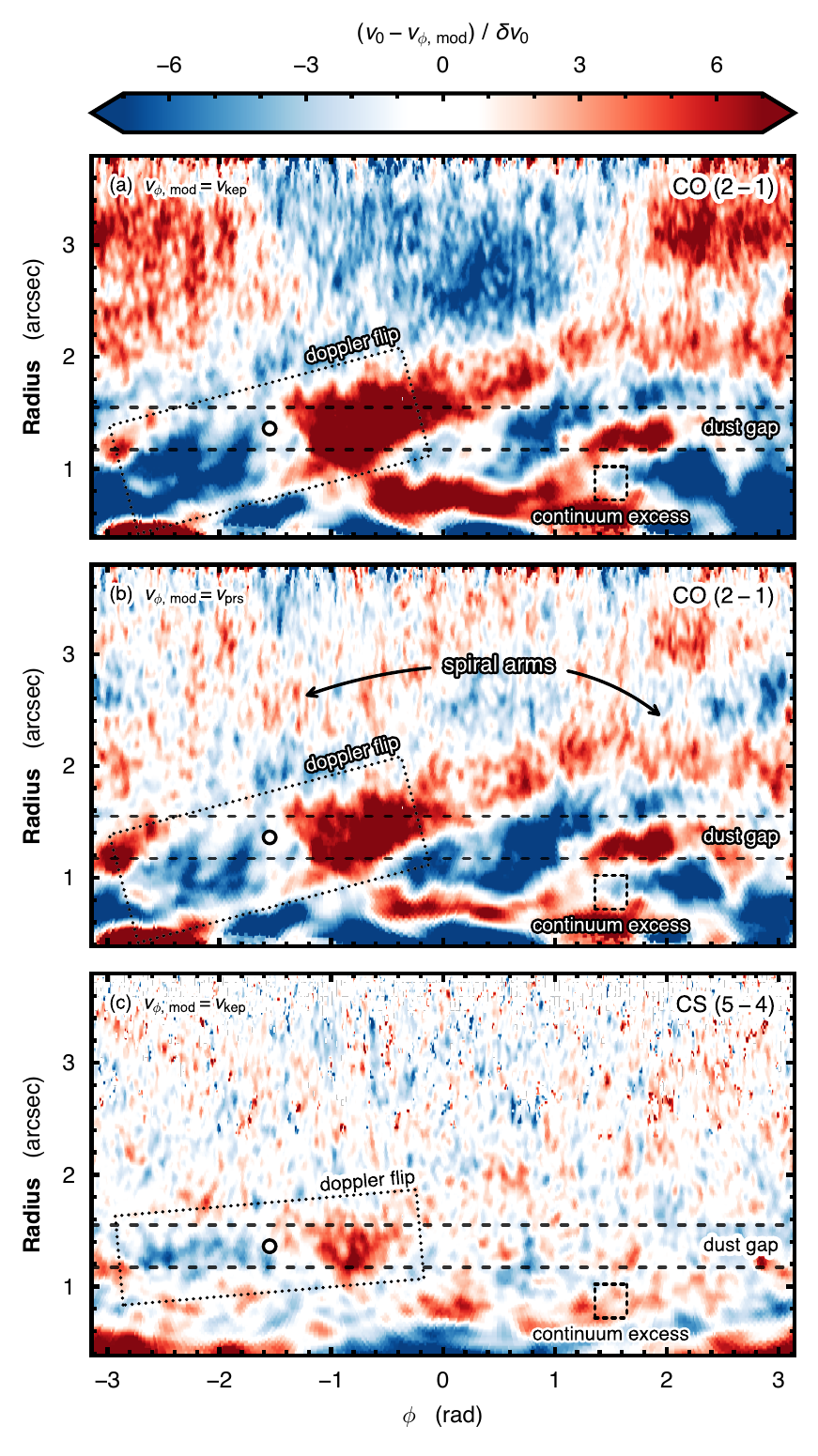}
    \caption{Polar projections of the velocity residual maps from Fig.~\ref{fig:v0_residuals} with CO after subtracting a purely Keplerian background profile in (a), and after subtracting a pressure-supported background in (b), and CS after subtracting a Keplerian background profile in (c). In all panels the residuals are normalized by the local uncertainty on the line center, $\delta v_0$. The two horizontal lines in each show the gap edges of the continuum gap at 82~au ($1\farcs35$) reported by \citet{Ilee_ea_2022}, coincident with the gap in scattered light \citep{Debes_ea_2013, Rapson_ea_2015, vanBoekel_ea_2017} and molecular line emission \citep{Teague_ea_2017}. The dotted boxes highlight the `Doppler flip' feature. The spirals traced by the CO are annotated. The circle marks the center of the `Doppler flip'. The dotted square marks the location of the continuum excess reporte in \citet{Tsukagoshi_ea_2019}. The disk rotates from right to left.}
    \label{fig:v0_residuals_polar}
\end{figure}

An alternative interpretation is to consider the case of when the underlying velocity structure is not constant in direction as a function of azimuth. For example, if there is in fact a `Doppler flip', a feature predicted to be indicative of an embedded planet \citep{Casassus_Perez_2019}, then a change in velocity direction would be expected at the location of the planet. \citet{Calcino_ea_2021}, following the analytical prescription of spiral wakes from \citet{Rafikov_2002}, demonstrate that the radial velocity components will be the dominant velocity component, except in the very immediate vicinity of the planet, where a radially inwards (negative $v_r$) component should be leading the planet, and a radially outwards (positive $v_r$) trailing the planet. This morphology is indeed what is seen for both CO and CS (Figures~\ref{fig:deprojected_v0_residuals}b and e), with the `Doppler flip' center aligning almost exactly with the disk minor axis. In this scenario, an embedded planet at a radial location of ${\sim}~1\farcs36$ (${\sim}~82~{\rm au}$) would naturally explain the presence of the surface density depletion at ${\sim}~85$~au, as well as providing a mechanisms for launching the large spiral structure which appears to emanate from a similar location. Figure~\ref{fig:v0_residuals_polar} demonstrates this proposed scenario, and is discussed more in Section~\ref{sec:discussion:planet}.

\begin{figure*}
    \centering
    \includegraphics[width=\textwidth]{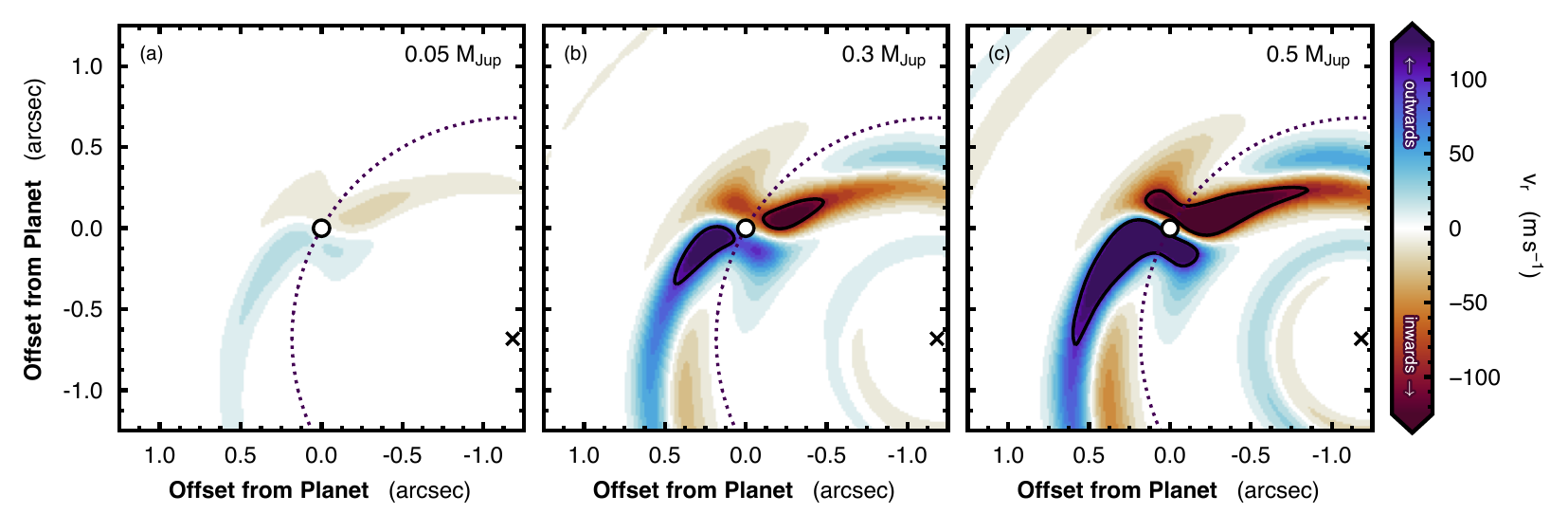}
    \caption{Radial velocity perturbations driven by planets of different masses: $1~M_{\rm Nep}$, (a), $1~M_{\rm Sat}$, (b), and $0.5~M_{\rm Jup}$, (c). Each panel is centered on the putative planet's location, with the dotted line showing the orbital radius of the planet, 82~au. The solid contours mark $100~{\rm m\,s^{-1}}$, the velocity traced by CS emission. The cross marks the center of the disk. Each panel spans a field of view of $2\farcs5$ (150~au). The analytical models were calculated following \citet{Bollati_ea_2021}.}
    \label{fig:analytic_models}
\end{figure*}

\section{Discussion}
\label{sec:discussion}

We have shown that the velocity structure of the protoplanetary disk around TW Hya, traced by CO and CS emission, exhibits substantial perturbations in the velocity structure. What is particularly interesting about this set of observations is that these molecules are believed to trace two distinct vertical regions in the disk: CO traces the disk atmosphere while CS traced closer to the disk midplane \citep[e.g.,][]{Dutrey_ea_2017}. This combination of lines therefore offer a glimpse of the vertical dependence of kinematic features within the disk. In this Section we discuss the implication of these findings.

\subsection{An Embedded Planet}
\label{sec:discussion:planet}

The search for planets in TW~Hya has had a long history. Multiple works have modeled various properties of the outer gap probed by different observations, be it total scattered light \citep{Debes_ea_2013}, polarized scattered light \citep{vanBoekel_ea_2017, Mentiplay_ea_2019}, dust continuum \citep{Ilee_ea_2022} or molecular line emission \citep{Teague_ea_2017}, to infer the mass of the putative planet responsible for opening the gap. These works have reported planet masses spanning between 10 -- $100~M_{\earth}$, all well below the current limits of direct detection methods which are sensitive down to a mass of ${\sim}~2~M_{\rm Jup}$ \citep{Asensio-Torres_ea_2021} at 90~au. The detection of localized velocity perturbations coincident with this gap therefore offers a new quantitative probe of the putative planet's mass.

\subsubsection{`Doppler Flip'}

The most intriguing aspect of the velocity residuals is the feature that bears a striking similarity to the `Doppler flip' morphology discussed in \citet{Casassus_Perez_2019}, where it was argued this it would be due to an embedded planet. This interpretation is particularly enticing for the case of TW~Hya due to coincidence of the `Doppler flip' and the outer gap location at ${\sim}~82$~au, as shown in Fig.~\ref{fig:v0_residuals_polar}. Such a coincidence between a localized velocity perturbation and a continuum gap has been previously argued to be a strong indication of an embedded planet \citep[e.g.,][]{Pinte_ea_2019}. Under the assumption that the motions associated with the `Doppler flip' are dominated by in-plane velocity components (i.e., either azimuthal or radial in direction) then Fig.~\ref{fig:deprojected_v0_residuals}b and e show that CO and CS emission are tracing perturbations on the order of $50~{\rm m\,s^{-1}}$ and $100~{\rm m\,s^{-1}}$, respectively. A stronger perturbation closer to the midplane is exactly what is predicted by numerical simulations as the spirals arising due to Lindblad resonances grow weaker with height in the disk \citep[e.g.,][]{Pinte_ea_2019}. Indeed, the variation in amplitudes of the perturbations found in the simulations of \citet{Pinte_ea_2019} are consistent with the difference in vertical heights expected to be traced by CO and CS emission. 

While a full hydrodynamical modeling of these observations are left for future work, the observed velocity perturbations can be compared with predictions from linear models \citep[e.g.,][]{Bollati_ea_2021} to estimate the putative planet's mass\footnote{\url{https://github.com/DanieleFasano/Analytical_Kinks}}. As these linear models only consider a flat, 2D disk, we limited this analysis to the CS emission for which the assumption of a 2D system is a more appropriate approximation than for the highly elevated region traced by CO emission. We assumed that the temperature and surface density of the disk over the region of interest are well described by single power laws with exponents $p = -0.75$ and $q = -0.47$, respectively \citep[e.g.,][]{Calahan_ea_2021, Huang_ea_2018}, which resulted in a local pressure scale height of $h = 0.073r$ at the radius of the planet when adopting the midplane temperature profile from \citet{Huang_ea_2018}.

Figure~\ref{fig:analytic_models} shows the $v_r$ components driven by an embedded planet with a mass of $0.05~M_{\rm Jup}$ ($1~M_{\rm Nep}$), $0.3~M_{\rm Jup}$ ($1~M_{\rm Sat}$) and $0.5~M_{\rm Jup}$, left to right, respectively, convolved with a Gaussian kernel with a FWHM of $0\farcs18$ to mimic the observations. Although the morphology of the perturbations are not a perfect match to the deprojected observations in Fig.~\ref{fig:deprojected_v0_residuals}, the magnitude of the perturbations, ${\sim}~100~{\rm m,\,s^{-1}}$, are best matched by the $0.3~M_{\rm Jup}$ planet, in line with the inferred planet mass based on the size of the gap. Perturbations in the azimuthal direction, $v_{\phi}$ are not shown as their morphology is inconsistent with the observed structures. However, we do note that as the $v_{\phi}$ motions for each planet mass are comparable in magnitude to the $v_r$ perturbations, regardless of the velocity component these observations are tracing, the magnitude of the perturbations are consistent with a Saturn-mass planet.

\subsubsection{Spiral}

A Saturn mass planet is also consistent with the sub-Jovian mass planet put forward in \citet{Bae_ea_2021} to account for the the large scale velocity spiral. As argued in Section~\ref{sec:decomposition:deprojection}, the azimuthal coherence of the spiral structure suggests that this must be dominated by vertical motions moving from the disk atmosphere towards the disk midplane. \citet{Bae_ea_2021} found that at large azimuthal distances from the planet, Lindblad spirals that drive the `Doppler flip' described above have negligible vertical motions, while buoyancy spirals were dominated by such vertical motions.

Comparing the velocity residuals shown in Fig.~\ref{fig:v0_residuals} to the numerical simulations of \citet{Bae_ea_2021}, it is clear that buoyancy resonances can qualitatively explain the morphology of the observed residuals, namely the tightly wound spirals with substantial vertical motions. The magnitude of the residuals from the least massive model considered -- a $0.5~M_{\rm Jup}$ planet -- is larger than what is observed in TW~Hya (${\sim}~10~{\rm m\,s^{-1}}$ if the perturbations are believed to be in the vertical direction). As the perturbation scale linearly with planet mass, the observed kinematic perturbations are therefore consistent with the inference of a Saturn mass planet based on the `Doppler flip' described above. Furthermore, the strong vertical dependence on the strength of the buoyancy spirals naturally explains the lack of a spiral in the CS data. As buoyancy resonances, the driving mechanism of the buoyancy spirals, require regions with slow thermal relaxation (i.e., a reduced rate of gas-particle collisions), they are expected to only grow in the atmospheric regions of disks, $z \, / \, r \gtrsim 0.1$. These regions would be therefore traced by CO emission, while the CS emission probes deeper in the disk \citep[$z\,/\,r \lesssim 0.1$;][]{Dutrey_ea_2017} where the thermal relaxation is sufficiently fast that buoyancy spirals cannot grow. 

A spiral in brightness temperature was reported in \citet{Teague_ea_2019a} through CO $(3-2)$ emission presented in \citet{Huang_ea_2018}. A direct comparison with the data presented here and this archival data is hindered by the detection of a large-scale shadow believed to be tracing the shadow observed in scattered light \citep{Debes_ea_2017, Teague_ea_2022}. Although \citet{Teague_ea_2022} did report the presence of a spiral arms in the outer disk consistent with that from \citet{Teague_ea_2019a}, a more comprehensive analysis would be limited by the different spectral response functions applied to each data set due to the differing spectral resolutions used when taking the data. Different spectral response functions will result in the intrinsic line profiles being modified in different ways leading to systematic differences in line widths or peak intensities that could be mistaken for true physical variations.

\subsubsection{Line Widths}

\begin{figure*}
    \centering
    \includegraphics[width=\textwidth]{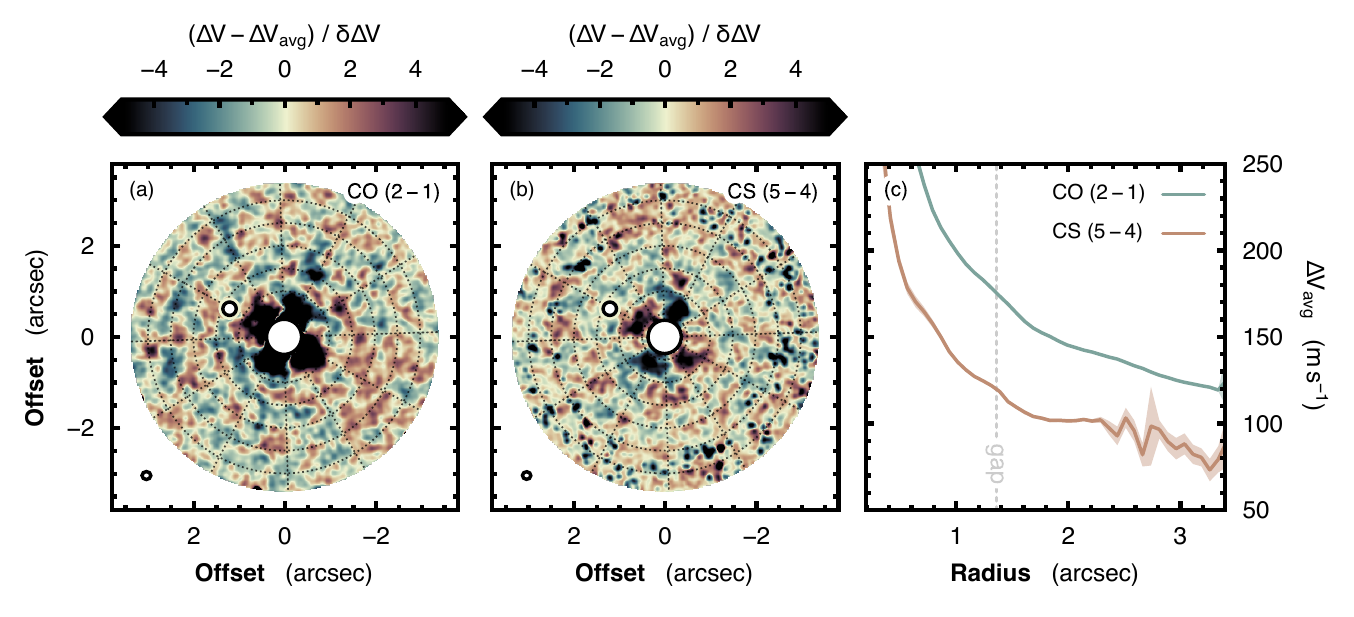}
    \caption{Residuals in the local line width, $\Delta V$, when an azimuthally averaged model, $\Delta\!V_{\rm avg}$, is subtracted for CO, (a), and CS, (b), normalized by the uncertainty of the linewidth measurement, $\delta\Delta\!V$. The radial profile used as the background model, $\Delta V_{\rm avg}$, is shown in (c). In both (a) and (b), the center of the proposed `Doppler flip' is shown by the white point, while the ellipse in the bottom left show the synthesized beam. Dotted contours show steps of $0\farcs5$ in radius and $30\degr$ in azimuth. In (c), the shaded regions shows teh $1\sigma$ standard deviation in each azimuthal bin. The vertical dotted line shows the location of the gap center as reported by \citet{Ilee_ea_2022}.}
    \label{fig:linewidth_plots}
\end{figure*}

Although this scenario paints a compelling picture of an embedded planet in TW~Hya, having the `Doppler flip' feature aligned with the disk minor axis limits the unambiguous decomposition of the velocity residuals. A search for enhanced line widths due to spatially unresolved motions around the putative embedded planet \citep[either from unresolved circumplanetary motion or planet-driven turbulent motions, e.g.,][]{Perez_ea_2015, Dong_ea_2019} would offer a method to potentially disentangle these scenarios. Unfortunately no evidence for localized enhancements were found in the maps of $\Delta V$ traced by CO or CS, as shown in Fig.~\ref{fig:linewidth_plots}.

A tentative enhancement in the background $\Delta V$ is found around the gap location, highlighted by the vertical dashed line in Fig.~\ref{fig:linewidth_plots}c. This is counter to the reduced line widths found in both $^{12}$CO and $^{13}$CO in the gaps of HD~163296 \citep{Izquierdo_ea_2022}, which were explained as due to the drop in local density limiting the pressure broadening of the line. One potential solution to this discrepancy is that for the face-on orientation of TW~Hya, observations are far more sensitive to the vertical velocity dispersions within the gap driven by the embedded planet \citep[e.g.,][]{Dong_ea_2019}. For a moderately inclined disk like HD~163296, the line of sight will only intersect a narrow region of the gap, tracing a smaller column of material, and thus a smaller velocity dispersion, than a line-of-sight directly into the gap.

A precise measurement of the width of a line requires substantially better data than what is needed for a similar precision on the line center \citep{Teague_2019}. As such, the most direct way to search for these unresolved motions would be to move to higher frequency observations, for example those tracing the $J=6-5$ transitions of CO isotopologues which can achieve far better velocity resolutions, improving on those reported here by almost a factor of 4. However, the larger upper state energies of these transitions may result in emission not being observable to the outer edge of the disk \citep[as was found for $^{13}$CO and C$^{18}$O emission][]{Schwarz_ea_2016}.

\subsection{Emission Heights}
\label{sec:discussion:emission_heights}

A particularly interesting point is that a similar dynamical stellar mass was found for both CO and CS: $0.814~M_{\odot}$ and $0.816~M_{\odot}$, respectively. As discussed in \citet{Yu_ea_2021}, if emission is expected to trace an elevated region parameterized by $\chi = z \, / \, r$, and a 2D Keplerian model is used to infer a stellar mass, then the true stellar mass would be underestimated by a factor of $(1 + \chi^2)^{3/2}$. If two different molecules are used to measure a dynamical mass, then the ratio of the inferred dynamical masses, $M_{\star,\,a}$ and $M_{\star,\,b}$ can be used to relate their two emission surfaces, $\chi_a$ and $\chi_b$, through,
\begin{equation}
    \chi_b = \left[ \left(1 + \chi_a^2 \right)\left( \frac{M_{\star,\,a}}{M_{\star,\,b}} \right)^{2/3} - 1 \right]^{1/2}.
\end{equation}
For the two stellar masses measured for CO and CS, this relationship would suggest CS traces a similar height to that of CO. This is seemingly at odds with what is observed in the Flying Saucer \citep{Dutrey_ea_2017} and found more generally from the MAPS program \citep{Law_ea_2021b}, where CO is observed to trace atmospheric disk regions, $\chi_{\rm CO}~{\sim}~0.3$, while CS is believed to trace closer to the disk midplane, $\chi_{\rm CS} \lesssim 0.1$. To add to this, \citet{Teague_ea_2018b} used multi-band observations of CS in TW~Hya to constrain the temperature of the gas, finding values close to 40~K in the inner 1\arcsec{} of the disk, and dropping to $\approx 20$~K beyond 2\arcsec{}. This is considerably lower than the temperature traced by CO emission by between 10 or 20~K \citep[e.g.,][]{Huang_ea_2018, Calahan_ea_2021}, suggesting that CO traces a more elevated, and thus hotter, region. Assuming then that CO and CS do trace different vertical heights, $\chi_{\rm CO} = 0.3$ and $\chi_{\rm CS} = 0.1$ for example, then the two dynamical masses should vary by 10\%. It is therefore likely that the large spiral structure observed in CO biases the inferred dynamical mass by roughly a similar level.

\subsection{CO Optical Depth}

\begin{figure}
    \centering
    \includegraphics[width=\columnwidth]{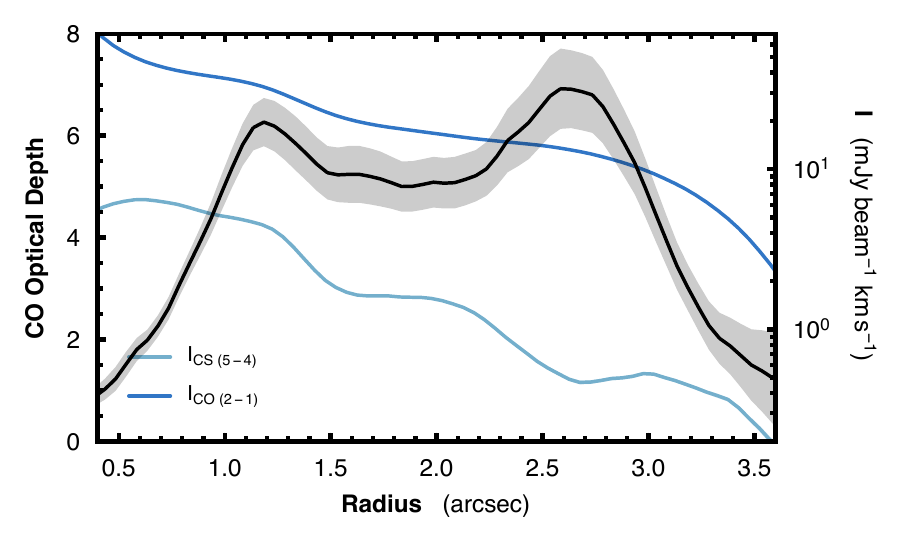}
    \caption{Comparing the azimuthally averaged radial profile of the CO optical depth, $\tau_0$ in black, with the azimuthally averaged radial profiles of the integrated intensities of the CO and CS emission, dark and light blue, respectively. The uncertainty shown on $\tau_0$ is dominated by the statistical uncertainty associated with the fitting of the line profile rather than azimuthal scatter.}
    \label{fig:CO_optical_depth}
\end{figure}

The arcs and spurs that are annotated in Fig.~\ref{fig:channels_CO_40ms} are found to be coincident in radius with the two rings of elevated optical depth at $1\farcs2$ (70~au) and $2\farcs65$ (160~au), as demonstrated in Fig.~\ref{fig:CO_optical_depth}. This suggests that at these radii there is a considerable increase in the local CO column density, enhancing the emission in the line wings which result in the observed protrusions which are symmetric about the line center.

What is particularly interesting is that the radial profile of $\tau_0$ exhibits the two aforementioned peaks separated by a broad plateau. This appears to bear little resemblance to the radial profile of the CO integrated intensity, shown as the dark blue line in Fig.~\ref{fig:CO_optical_depth}, which appears relatively smooth. However, a subtle bump is seen at $1\farcs2$, as previously reported by \citet{Huang_ea_2018}, consistent with a local enhancement in the CO column density. Comparison with the radial profile of $^{13}$CO \citep[Fig.~7 of][]{Zhang_ea_2019}, which is less optically thick than $^{12}$CO, does show peaks in integrated intensity at broadly the same radial locations at peaks in $\tau_0$, further evidencing that these peaks are indeed tracing annuli of enhanced CO column density. These observations therefore demonstrate that with sufficient spectral resolution the subtle change in emission line morphology -- namely the saturation of the core -- enables a characterization of column density variations even in highly optically thick lines.

The comparison with the CS integrated intensity, shown in light blue in Fig.~\ref{fig:CO_optical_depth}, however, is more complex. The inner peak at $1\farcs2$ (70~au) does appear to align with the outer edge of the CS ring discussed in Section~\ref{sec:observations:moment_maps}, signaling a localized enhancement in the total gas surface density rather than just CO column density. This is also consistent with the rings observed in CN emission \citep{Teague_Loomis_2020} and DCN emission \citep{Oberg_ea_2021} which align with the edge of the pebble disk. The outer peak in $\tau_0$ at $2\farcs65$ (160~au), however, instead aligns with a drop in both CS and CN integrated intensity. \citet{Teague_ea_2018c} used a multi-transition excitation analysis to demonstrate that this emission deficit was due to a drop in CS column density rather than a change in temperature. Similarly \citet{Teague_Loomis_2020} showed that no change in the excitation temperature of CN was found, requiring instead a drop in CN column density to account for the dip. The difference in the behavior of the CO emission to CS and CN emission therefore points towards chemical processing that could lead to a preferential increase in CO column densities relative to other species. A more thorough exploration of chemical scenarios that can lead to this difference is left for future work.

\section{Summary}
\label{sec:summary}

We have presented new, high velocity resolution observations of CO and CS emission from the disk around TW~Hya. We have used these to map and characterize the complex kinematic structure of the disk. Line-of-sight velocity maps were calculated by fitting analytical forms to the spectrum in each pixel and the bulk background motion was found to be well described by a Keplerian profile. Regions beyond $2\farcs1$ (126~au) traced by CO emission were found to be rotating at slightly sub-Keplerian speeds, likely due to the influence of a strong radial pressure gradient at the edge of the disk \citep{Dullemond_ea_2020}.

After subtracting the best-fit rotation model from the data, complex kinematic substructure was found in the gas velocities traced by both molecules, most notably a large, open spiral traced by CO, and a `Doppler flip' feature traced by both CO and CS. The location of the `Doppler flip`, at a radial offset from the central star of $1\farcs35$ (82~au) and position angle ${\rm PA} = 60\degr$, lies along the north-east minor axis of the disk and coincides with a gap in the disk. This coincidence in location is strong evidence of an embedded planet opening the gap \citep{Pinte_ea_2019}.

A Saturn-mass planet ($0.3~M_{\rm Jup}$) was found to reproduce the magnitude of the velocity perturbations associated with the `Doppler flip', while also falling within the range of masses inferred based on the width and depth of the outer gap \citep[e.g.,][]{Debes_ea_2013, Teague_ea_2017, vanBoekel_ea_2017, Mentiplay_ea_2019, Ilee_ea_2022}. Furthermore, the modeling of \citet{Bae_ea_2021} showed that a sub-Jovian mass planet is capable of launching buoyancy spirals in the disk atmosphere, which exhibit a similar morphology and magnitude to those traced by the CO emission. The interpretation of this spiral as due to buoyancy resonances, which are found only in the atmospheres of disks, naturally explains why it is traced by CO emission but not CS emission.

Although no elevated velocity dispersions were found around the `Doppler flip' location, a tentative enhancement in the background line width was found at the radius of the gap. Such motions could be interpreted as highly localized vertical motions excited by the embedded planet within the gap \citep[e.g.,][]{Dong_ea_2019}.

In sum, these results show compelling evidence of an embedded, Saturn-mass planet in the disk around TW~Hya, carving out the gap at 82~au ($1\farcs35$). They demonstrate the power of observations designed to trace the kinematical properties of disks, while also emphasizing the utility of coupling molecules known to trace distinct vertical regions in order to extract critical information about the 3D structure of the features found.

\section*{Acknowledgments}
We thank the anonymous referee for their thorough reading of the manuscript and their helpful comments. This paper makes use of the following ALMA data: ADS/JAO.ALMA\#2013.1.00387.S and ADS/JAO.ALMA\#2018.A.00021.S. ALMA is a partnership of ESO (representing its member states), NSF (USA) and NINS (Japan), together with NRC (Canada), MOST and ASIAA (Taiwan), and KASI (Republic of Korea), in cooperation with the Republic of Chile. The Joint ALMA Observatory is operated by ESO, AUI/NRAO and NAOJ. The National Radio Astronomy Observatory is a facility of the National Science Foundation operated under cooperative agreement by Associated Universities, Inc. Based on observations made with the NASA/ESA Hubble Space Telescope, obtained from the Data Archive at the Space Telescope Science Institute, which is operated by the Association of Universities for Research in Astronomy, Inc., under NASA contract NAS5-26555. These observations are associated with program \#16228. This project has received funding from the European Research Council (ERC) under the European Union’s Horizon 2020 research and innovation programme (grant agreement No. 101002188) Support for J. H. was provided by NASA through the NASA Hubble Fellowship grant \#HST-HF2-51460.001-A awarded by the Space Telescope Science Institute, which is operated by the Association of Universities for Research in Astronomy, Inc., for NASA, under contract NAS5-26555. 

\software{emcee \citep{emcee}, GoFish \citep{GoFish}, bettermoments \citep{Teague_Foreman-Mackey_2018}, CASA \citep[v5.8.0;][]{casa}, \texttt{keplerian\_mask.py} \citep{keplerian_mask}}

\appendix

\section{Moment Maps}
\label{sec:app:moments}

In this Appendix we present the moment maps made from the `optically thick Gaussian' and Gaussian analytical fits discussed in Section~\ref{sec:observations:moment_maps} for CO and CS, respectively. The maps were generated with the \texttt{bettermoments} package. We note that although the optical depth is accounted for with the $\tau_0$ variable, the spectral response function of the ALMA correlator lessens the strength of the optically thick core due to the convolution of the signal with the spectral response function, resulting in an underestimation of the true optical depth. 

\begin{figure}
    \centering
    \includegraphics[width=\textwidth]{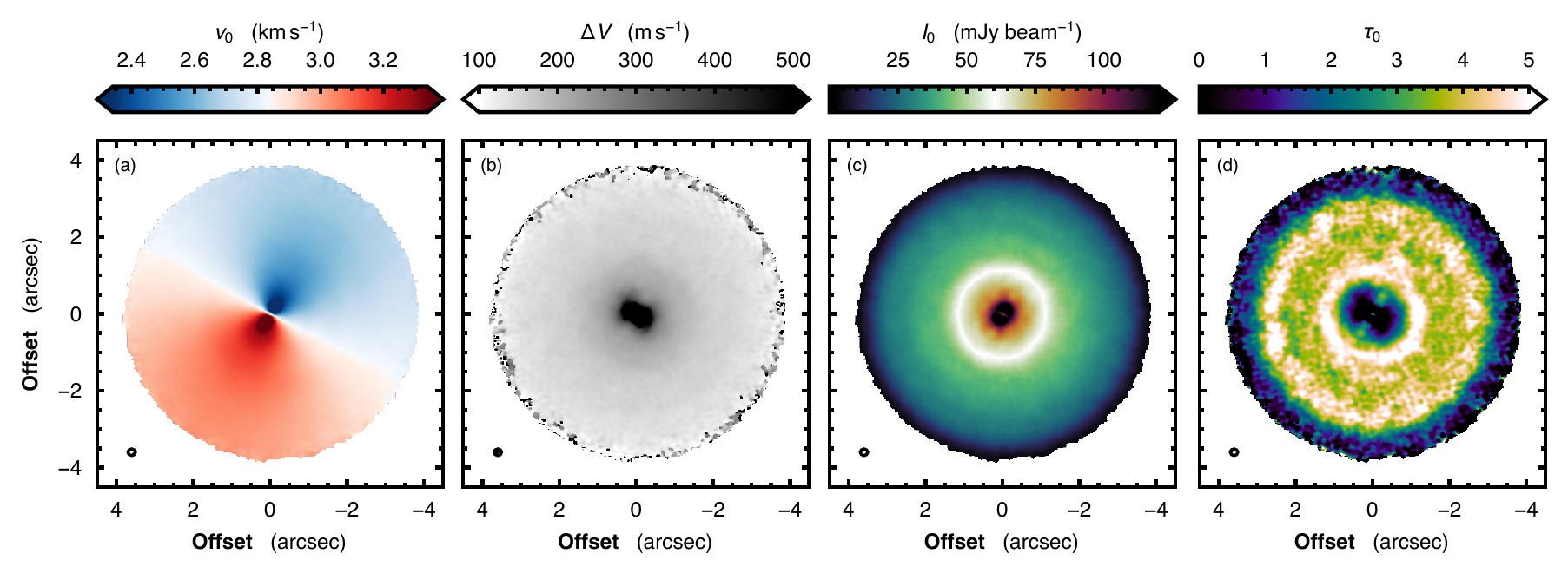}
    \caption{Moment maps of the CO emission. Only regions where $I_0 > 5\sigma$, where $\sigma = 0.7~{\rm mJy~beam^{-1}}$, are shown.}
    \label{fig:app:CO_moment_maps}
\end{figure}

\begin{figure}
    \centering
    \includegraphics{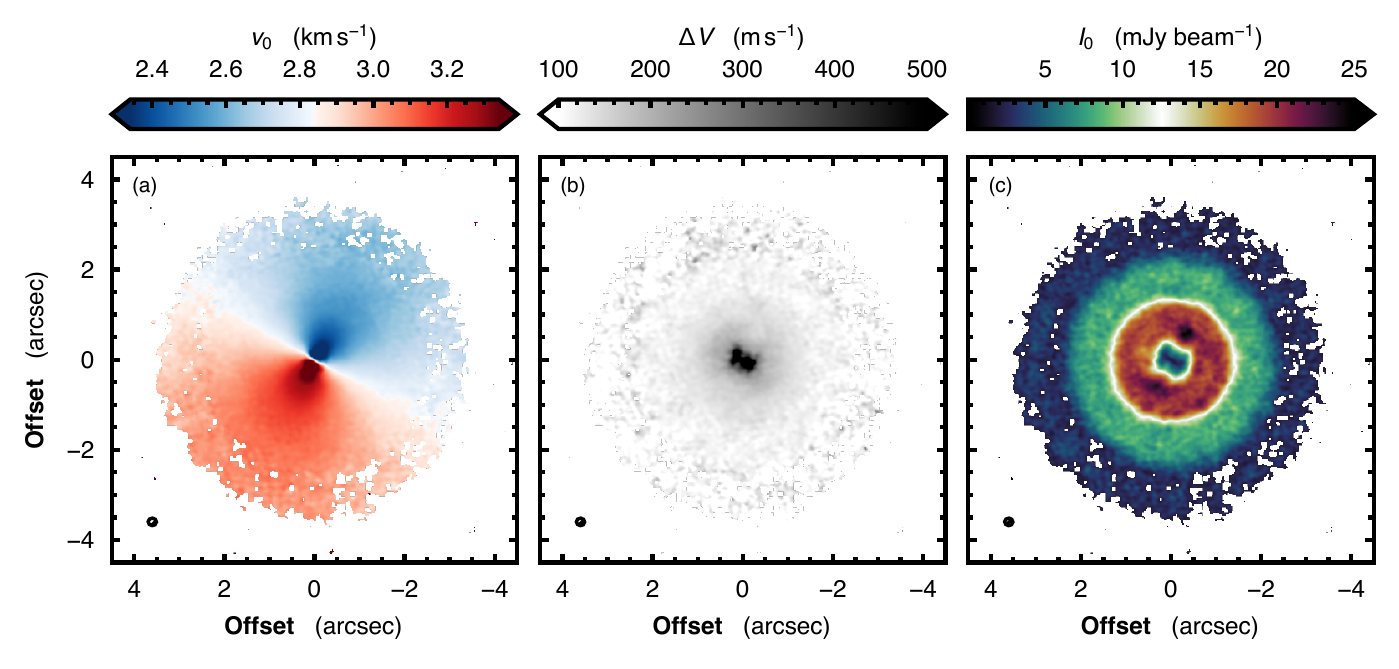}
    \caption{Moment maps of the CS emission. The $v_0$ and $\Delta V$ panels are shown on the same scaling as CO in Fig.~\ref{fig:app:CO_moment_maps}. Only regions where $I_0 > 3\sigma$, where $\sigma = 0.3~{\rm mJy~beam^{-1}}$, are shown.}
    \label{fig:app:CS_moment_maps}
\end{figure}

\section{Theoretical Limits in Extracting Velocity Perturbations}
\label{sec:app:accuracy}

It is useful to consider what the theoretical limits are in terms of detecting velocity deviations as to know what deviations we should be able to detect in a given data set. In practice, this is answering the question \emph{``How well can we measure the line centroid?''}. In this Appendix, we build upon the work of \citet{Lenz_Ayres_1992} who were able to define a relationship between the sampling rate, $R$, defined as the number of samples over the FWHM of the line, and the signal-to-noise ratio of the spectrum peak, SNR, to the accuracy to which parameters describing a Gaussian profile could be constrained. From this relationship, it was possible to estimate what level of velocity deviation one could expect to detect for a data set.

\begin{figure}
    \centering
    \includegraphics{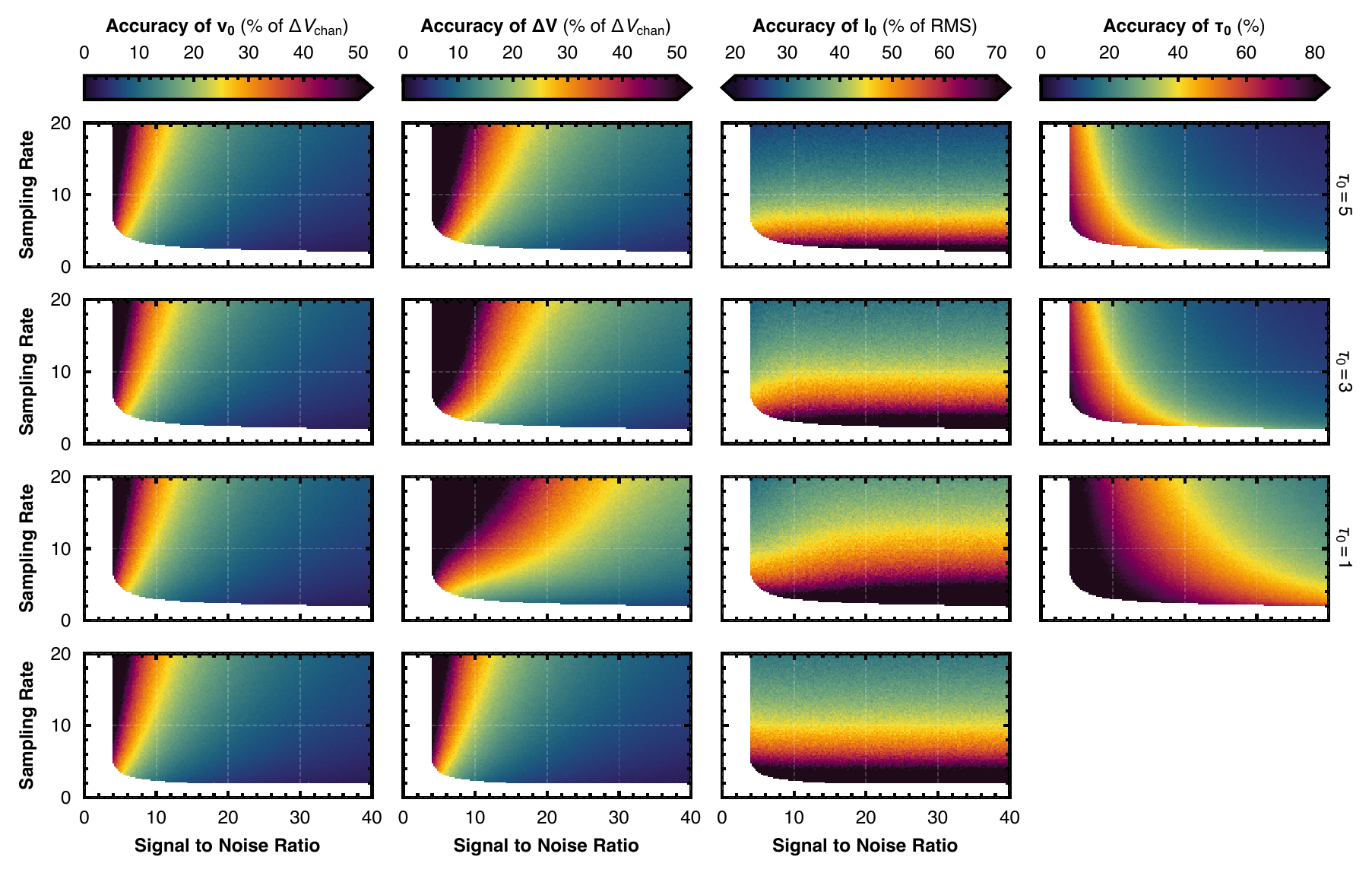}
    \caption{The accuracy to which the parameters of a Gaussian profile can be fit to noisy data for a given sampling rate, $R$, and signal-to-noise ratio, SNR. Each column represents a different parameter, each row represents a different $\tau_0$: 5, 3, 1 and 0, top to bottom.}
    \label{fig:recovery_of_parameters}
\end{figure}

Here, we extend this methodology to an opacity-broadened Gaussian profile, more suited to the optically thick $^{12}$CO emission presented in this paper. The methodology is as follows. Firstly, we assume the underlying true spectrum is described by,
\begin{equation}
    I(v) = I_0 \cdot \frac{1 - \exp\big(-\tau(v)\big)}{1 - \exp(-\tau_0)} ,
\end{equation}
where $I_0$ and $\tau_0$ are the intensity and optical depth at the line center, and where $\tau(v)$ has a Gaussian form,
\begin{equation}
    \tau(v) = \tau_0 \exp \left( -\left[ \frac{v - v_0}{\Delta V} \right]^2 \right),
\end{equation}
where $v_0$ is the line center and $\Delta V$ is the Doppler width (\emph{not} the typical standard deviation of a Gaussian, but rather a factor of $\sqrt{2}$ larger, such that ${\rm FWHM} = 2 \sqrt{{\rm ln}2}~\Delta V$). Then, for a given $\{R,\, {\rm SNR},\, \tau_0\}$ set, a model spectrum is generated on a $v$-axis sampled at integer values, with the line center, $v_0$, randomly drawn from a Normal distribution. This defines the `true' spectrum. Gaussian noise is then added to the model spectrum, with a standard deviation of $\sigma = I_0 \, / \, {\rm SNR}$, creating the `observed' spectrum. The assumption of white noise is not perfect, however is a reasonable approximation for ALMA data that has been imaged with a channel spacing larger than the spectral resolution.

Using the Python package \texttt{lmfit}, a model profile is fit to the `observed' spectrum by minimizing $\chi^2$ using the Levenberg-Marquardt algorithm. This returns optimized model parameters and their associated uncertainties (while the full covariance matrix is returned, we only take the diagonal components and ignore any parameter covariances for this analysis). The difference between the optimized parameter and the true parameter is recorded as a measure of the accuracy of the fit. This process is then repeated $N$ times, with each iteration using a new, random spectrum. It was found that for $N \gtrsim 10^3$, the distribution of (true - optimized) for a given parameter followed a Gaussian distribution, with a mean equal to the average uncertainty returned from the fitting. To ensure this distribution was well sampled, we adopted $N = 10^4$ for any $\{R,\, {\rm SNR},\, \tau_0\}$ set.

We then sample the $\{R,\, {\rm SNR},\, \tau_0\}$ parameter space. Four values of $\tau_0$ were chosen: 0 (a pure Gaussian model to compare with \citealt{Lenz_Ayres_1992}), 1, 3 and 5. For $\tau_0 > 5$ the line is almost entirely saturated and very little difference in the resulting spectra can be seen. The sampling rate, $R$, was chosen to span from 2 to 20, covering the range of possible rates based on typical line widths (between \ms{100} and \ms{250}, the widths of thermally broadened lines at temperatures between 20~K and 100~K) and spectral resolutions afforded by ALMA (roughly between \ms{30} and \ms{250}). While many observations result in $R < 2$, there is little utility in fitting an analytical model to data so poorly sampled, and thus these values are ignored. The SNR was chosen to range between 3 and 40, again typical of ALMA observations of molecular line emission from protoplanetary disks.

The standard deviation of the (true - optimized) distribution for each parameter and $\{R,\, {\rm SNR},\, \tau_0\}$ set is shown in Fig.~\ref{fig:recovery_of_parameters}. Both the line center, $v_0$, first column, and the line width, $\Delta V$, second column, are normalized relative to the sampling rate. As can clearly be seen, and as expected, the accuracy increases with both an increased $R$ and SNR. Adopting a similar functional form to that used in \citet{Lenz_Ayres_1992}, we can model the relationship between the accuracy to which the line center can be determined and the data characteristics. Again using \texttt{lmfit}, we find $\delta(v_0) \, / \, \Delta V_{\rm chan} = 0.69 \,\sqrt{R} \, / \, {\rm SNR}$ when averaged over all four $\tau_0$ values. Applying this relationship to the data presented here, the channel spacing is \ms{40}. The typical Doppler widths of the lines are \ms{160}, resulting in $R \approx 6.7$. Assuming that the ${\rm SNR} \sim 10$ for a given pixel, then $\delta(v_0) \approx$~\ms{7}, or about a fifth of a channel. For the other three parameters, there is a strong $\tau_0$ dependency which limits the utility in defining a similar relationship.

\section{Beyond Keplerian Rotation}
\label{sec:app:beyond_keplerian}

With improvements in both the quality of observations and the analysis techniques, it is no longer reasonable to assume a purely Keplerian background rotation profile as the contribution from either the disk self-gravity \citep[e.g.,][]{Veronesi_ea_2021}, or the background pressure gradient \citep[e.g.,][]{Dullemond_ea_2020} are now discernible. As discussed in Section~\ref{sec:decomposition:background}, methods that split the observations into independent annuli can circumvent explicitly dealing with these contributions as the azimuthally average $v_{\phi}$ profile can be easily reconstructed by interpolating between each annulus. However, as also discussed in Section~\ref{sec:decomposition:background}, there is still utility in developing a parameterized profile for situations where strong azimuthal variations in $v_0$ prevent the annulus-by-annulus approach. In this Appendix we discuss a parameterization which was included in the \texttt{eddy} code \citep{eddy}, originally developed to search for the impact of disk self-gravity on the rotation curve, but was later found to provide a reasonable description of the sub-Keplerian rotation due to the radial pressure gradient.

\begin{figure}
    \centering
    \includegraphics[width=\textwidth]{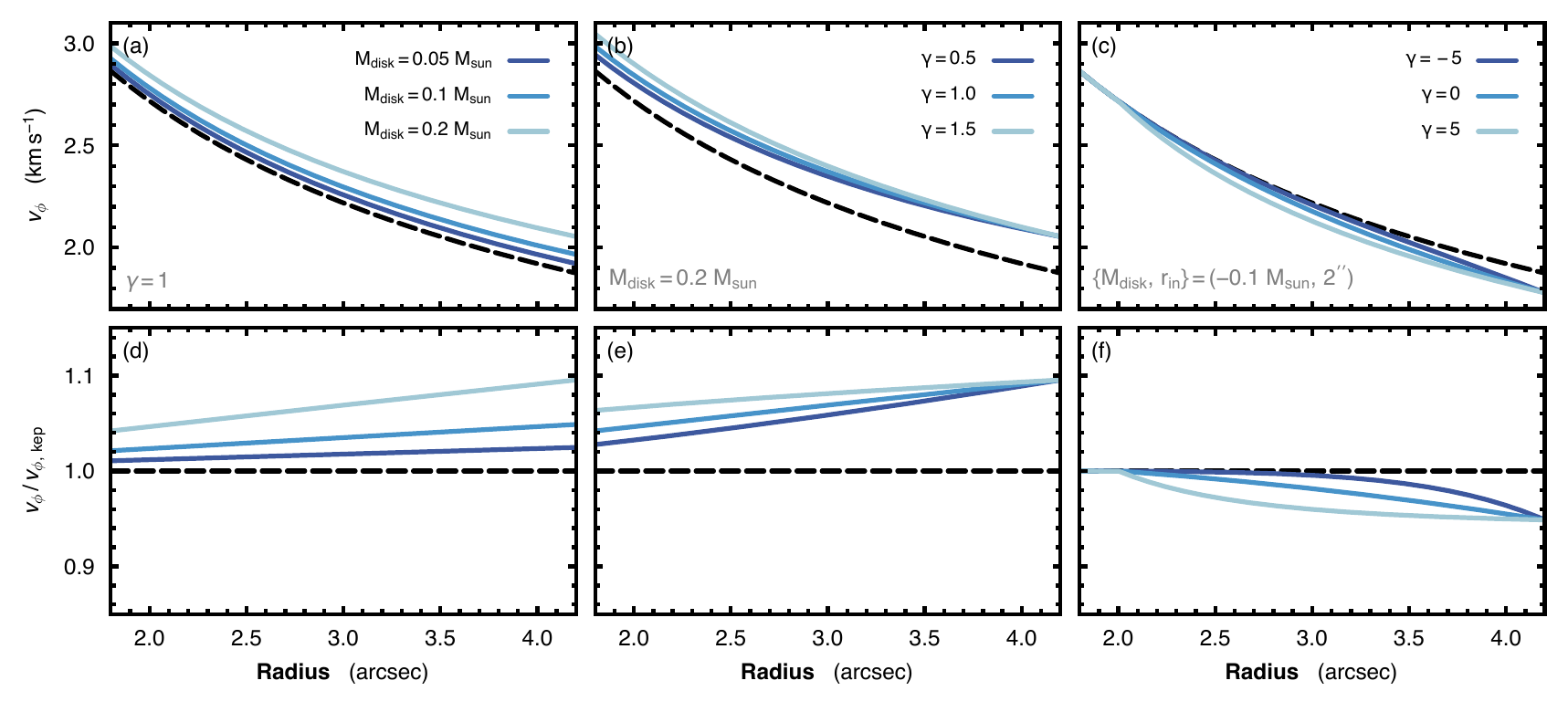}
    \caption{Rotation profiles using the new parameterization to model non-Keplerian components. The top row shows the $v_{\phi}$ in ${\rm km\,s^{-1}}$, while the bottom row shows them relative to a purely Keplerian rotation profile. Panels (a) and (d) show the impact of changing $M_{\rm disk}$ while panels (b) and (e) shows the impact of changing $\gamma$. Panels (c) and (f) demonstrate how a negative $M_{\rm disk}$ in this parameterization can be used to model sub-Keplerian rotation in the outer disk.}
    \label{fig:app:analytical_deviations}
\end{figure}

To mimic the effect of disk self-gravity on the rotational velocity profile we have included a disk mass term, $M_{\rm disk}$, such that the total dynamical mass of the system is given by $M_{\rm tot} = M_{*} + f_d(r) \cdot M_{\rm disk}$, where $f_d(r)$ describes the fraction of the disk mass with within a cylindrical radius $r$. Assuming that the disk surface density is described by a power law profile, $\Sigma(r) = \Sigma_0 \, (r\, / \,r_c)^{-\gamma}$, then the disk mass interior to $r$ is given by
\begin{equation}
    f_d(r) = \frac{r^{2 - \gamma} - r_{\rm in}^{2 - \gamma}}{r_{\rm out}^{2 - \gamma} - r_{\rm in}^{2 - \gamma}},
\end{equation}
where $r_{\rm in}$ and $r_{\rm out}$ describe the inner and outer disk edges. For $\gamma = 1$, this is simply a linear interpolation between 0 and 1 from $r_{\rm in}$ to $r_{\rm out}$. Panels a and b in Figure~\ref{fig:app:analytical_deviations} demonstrate the effect of changing $M_{\rm disk}$ and $\gamma$ on $v_{\phi}$, respectively.

While this parameterization is not as accurate as the full form presented in \citet[notably it doesn't take into account mass \emph{beyond} the cylindrical radius $r$]{Veronesi_ea_2021}, it allows for a simple way to mimic the sub-Keplerian rotation due to pressure support in the outer disk by considering negative values of $M_{\rm disk}$. In this form, shown in panels c and f in Figure~\ref{fig:app:analytical_deviations}, $M_{\rm disk}$ and $\gamma$ can be used to introduce sub-Keplerian motion in regions beyond $r_{\rm in}$. Although a direct relation cannot be made to the true pressure support term,
\begin{equation}
    v_{\rm prs}^{2} = \frac{r}{\rho}\frac{\partial P}{\partial r},
\end{equation}
the value of $r_{\rm in}$ can be broadly interpreted as where the exponential tail of the gas surface density kicks in, while $\gamma$ can be used to quantify how rapidly the rotation curve deviates from Keplerian. For a full exploration of the impact of pressure support, full forward modeling is required \citep[e.g.,][]{Dullemond_ea_2020, Yu_ea_2021}.

\bibliography{main}
\bibliographystyle{aasjournal}

\end{document}